\documentclass[twocolumn,letterpaper,amsmath,amssymb]{revtex4}
\usepackage{graphicx}
\usepackage{dcolumn}
\usepackage{bm}
\begin{document}

\title{ Edwards thermodynamics of the jamming transition for
  frictionless packings: ergodicity test and role of angoricity and
  compactivity}

\author{Kun Wang$^{1}$, Chaoming Song$^{2}$, Ping Wang$^{3}$, Hern\'an A. Makse$^{1}$}

\affiliation {$^{1}$ Levich Institute and Physics Department, City College of New York, New York, NY 10031, US \\
  $^{2}$ Center for Complex Network Research, Department of Physics, Biology and Computer Science, Northeastern University, Boston, MA 02115, US \\
  $^{3}$ FAS Center for Systems Biology, Harvard University,
  Cambridge, MA 02138, US}

\date{\today }

\begin{abstract}

  This paper illustrates how the tools of equilibrium statistical
  mechanics can help to explain a far-from-equilibrium problem: the
  jamming transition in frictionless granular materials. Edwards’
  ideas consist of proposing a statistical ensemble of volume and
  stress fluctuations through the thermodynamic notion of entropy,
  compactivity, $X$, and angoricity, $A$ (two temperature-like
  variables). We find that Edwards’ thermodynamics is able to describe
  the jamming transition (J-point).  Using the ensemble formalism we
  elucidate the following: {\it (i)} We test the combined
  volume-stress ensemble by comparing the statistical properties of
  jammed configurations obtained by dynamics with those averaged over
  the ensemble of minima in the potential energy landscape as a test
  of ergodicity.  Agreement between both methods supports the idea of
  ``thermalization'' at a given angoricity and compactivity.  {\it
    (ii)} A microcanonical ensemble analysis supports the idea of
  maximum entropy principle for grains. {\it (iii)} The intensive
  variables describe the approach to jamming through a series of
  scaling relations as $A\to 0^+$ and $X\to 0^-$. Due to the
  force-volume coupling, the jamming transition can be probed
  thermodynamically by a ``jamming temperature'' $T_{\rm J}$ comprised
  of contributions from $A$ and $X$.  {\it (iv)} The thermodynamic
  framework reveals the order of the jamming phase transition by
  showing the absence of critical fluctuations at jamming in
  observables like pressure and volume.
  {\it (v)} Finally, we elaborate on a comparison with relevant
  studies showing a breakdown of equiprobability of microstates.

\end{abstract}
\maketitle


The application of concepts from equilibrium statistical mechanics to
out of equilibrium systems has a long history of describing diverse
systems ranging from glasses to granular materials
\cite{coniglio,behringer,edwardsbook1}. For dissipative jammed
systems--- particulate grains or droplets--- the key concept proposed
by Edwards is to replace the energy ensemble describing conservative
systems by the volume ensemble \cite{edwardsbook1}.  However, this
approach alone is not able to describe the jamming point
(J-point) for deformable particles like emulsions and droplets
\cite{notcool,jpoint,powerlaw,mode2}, whose geometric configurations
are influenced by the applied external stress.  Therefore, the volume
ensemble requires augmentation by the ensemble of stresses
\cite{angoricity,forcebalance1,forcechain,forcemap}. Just as volume
fluctuations can be described by compactivity, the stress fluctuations
give rise to an angoricity, another analogue of temperature in
equilibrium systems.

In the past 20 years since the publication of Edwards’ work there has
been many attempts to understand and test the foundations of the
thermodynamics of powders and grains.  Three approaches are relevant
to the present study:

\begin{itemize}

\item[\bf 1.] {\bf Experimental studies of reversibility.---} Starting
  with the experiments of Chicago which were reproduced by other
  groups \cite{chicago1,bideau,swinney,sdr}, a well-defined
  experimental protocol has been introduced to achieve reversible
  states in granular matter.  These experiments indicate that
  systematically shaken granular materials show reversible behavior
  and therefore are amenable to a statistical mechanics approach,
  despite the frictional and dissipative character of the
  material. These results are complemented by direct measurements of
  compactivity and effective temperatures in granular media
  \cite{chicago1,swinney,danna,song-pnas,wang-pre}.

\item[\bf 2.] {\bf Numerical test of ergodicity.---} Numerical
  simulations compare the ensemble average of observables with those
  obtained from direct dynamical measures in granular matter and
  glasses. These studies \cite{nicodemi,mod1,barrat,mod2,mk,ciamarra}
  find general agreement between both measures and, together with the
  experimental studies of reversibility
  \cite{chicago1,bideau,swinney,sdr}, suggest that ergodicity might
  work in granular media.

\item[\bf 3.] {\bf Numerical and experimental studies of
    equiprobability of jammed states.---} Exhaustive searches of all
  jammed states are conducted in small systems to test the
  equiprobability of jammed states, as a foundation of the
  microcanonical ensemble of grains. Numerical simulations and
  experiments indicate that jammed states are not equiprobable
  \cite{ohern,manypackings,shattuck}. These results suggest that a
  hidden extra variable \cite{leche} might be needed to describe
  jammed granular matter in contrast with the work in {\bf 1} and {\bf
    2}.

\end{itemize}

The current situation can be summarized as following: When directly
tested or exploited in practical applications, Edwards ensemble seems
to work well.  These include studies where ensemble and dynamical
measurements are directly compared, and recent applications of the
formalism to predict random close packing of monodisperse spherical
particles \cite{compactivity,briscoe}, polydisperse systems
\cite{max}, and two \cite{meyer} and high dimensional systems
\cite{yin}.  However, a direct count of microstates reveals problems
at the foundation of the framework manifested in the breakdown of the
flat average assumption in the microcanonical ensemble
\cite{ohern,manypackings,shattuck,leche}.

In this paper we investigate the Edwards ensemble of granular matter
focusing on describing the jamming transition
\cite{notcool,jpoint,powerlaw,mode2}.  A short version of this study
has been recently published in \cite{wang-epl}.  We employ a strategy
that mixes the approaches {\bf 2} and {\bf 3} above. We first perform
an exhaustive search of all jammed configurations in the Potential
Energy Landscape (PEL) of small frictionless systems in the spirit of
\cite{ohern,manypackings,shattuck,leche}. We then use this information
to perform a direct test of ergodicity in the spirit of
\cite{nicodemi,mod1,barrat,mod2,mk,ciamarra}.  Our results indicate:
{\it (i)} The dynamical and ensemble measurements of presure,
coordination number, volume, and distribution of forces agree well,
supporting ergodicity. A microcanonical ensemble analysis supports
also a maximum entropy principle for grains.  {\it (ii)} Intensive
variables like angoricity, $A$, and compactivity, $X$, describe the
approach to jamming through a series of scaling relations.  Due to the
force-volume coupling, the jamming transition can be probed
thermodynamically by a ``jamming temperature'' $T_{\rm J}$ comprised
of contributions from $A$ and $X$.  {\it (iii)} These intensive
variables elucidate the thermodynamic order of the jamming phase
transition by showing the absence of critical fluctuations above
jamming in static observables like pressure and volume.  That is, the
jamming transition is not critical and there is no critical
correlation length arising from a thermodynamic n-point correlation
function.  We discuss other possible correlation lengths.  {\it (iv)}
Surprisingly, we reproduce the results of \cite{ohern} regarding the
failure of equiprobability of microstates while obtaining the correct
dynamics measurements as in
\cite{nicodemi,mod1,barrat,mod2,mk,ciamarra}.  We then offer a
possible solution to this conundrum to elucidate why the microstates
seems to be not equiprobable while the ensemble averages produce the
correct results.

The paper is organized as follows. Section \ref{edw} discusses the
Edwards thermodynamics of the jamming transition.  Section \ref{pel-c}
describes the ensemble calculations in the Potential Energy Landscape
formalism.  Section \ref{hertz} describes the Hertzian system to be
studied.  Section \ref{dos} describes the ensemble measurements to be
compared with the MD measures of Section \ref{MD}.  Section
\ref{angoricity} explains how to calculate $A$ from the data.  The
ergodicity test is made in Section \ref{ergodicity}.  Section
\ref{thermo} describes the calculation in the microcanonical ensemble
where the principle of maximum entropy is verified and the coupled
jamming temperature is obtained.  Section \ref{comparison} compares
our results with those of O' Hern {\it et al.} \cite{ohern} and
Section \ref{conclusion} summarizes the work. Appendix \ref{z}
includes ``de yapa'' a study of coordination number fluctuations in
the Edwards theory for random close packings of hard spheres.

\section{Edwards thermodynamics and the jamming transition}
\label{edw}

The process typically referred to as the jamming transition occurs at
a critical volume fraction $\rm {\phi_c}$ where the granular system
compresses into a mechanically stable configuration in response to the
application of an external strain \cite{coniglio,behringer,notcool}.
The application of a subsequent external pressure with the concomitant
particle rearrangements and compression results in a set of
configurations characterized by the system volume $V=N V_g/\phi$
($\phi$ is the volume fraction of $N$ particles of volume $V_g$) and
applied external stress or pressure $p$ (for simplicity we assume
isotropic states).

It has been long argued whether the jamming transition is a
first-order transition at the discontinuity in the average
coordination number, $Z$, or a second-order transition with the
power-law scaling of the system's pressure as the system approaches
jamming with $\phi-\phi_c \to 0^+$
\cite{jpoint,powerlaw,hmlaw2,mode2}. Previous work
\cite{forcemap,gama,canonicaljam} has proposed to explain the jamming
transition by a field theory in the pressure ensemble. Here, we use
the idea of ``thermalization'' of an ensemble of mechanically stable
granular materials at a given volume and pressure to study the jamming
transition from a thermodynamic viewpoint.

For a fixed number of grains, there exist many jammed states
\cite{ohern,manypackings} confined by the external pressure $p$ in a
volume $V$. In an effort to describe the nature of this nonequilibrium
system from a statistical mechanics perspective, a statistical
ensemble \cite{angoricity,forcechain,forcemap} was introduced for
jammed matter. In the canonical ensemble of pressure and volume, the
probability of a state is given by $\exp [-{\cal W} (\partial
S/\partial V) - \Gamma (\partial S/\partial \Gamma)]$, where $S$ is
the entropy of the system, $\cal W$ is the volume function measuring
the volume of the system as a function of the particle coordinates and
$\Gamma \equiv pV$ is the boundary stress (or internal virial)
\cite{gama} of the system. Just as $\partial E/\partial S = T$ is the
temperature in equilibrium system, the temperature-like variables in
jammed systems are the compactivity \cite{edwardsbook1}
\begin{equation}
  X=\partial V/\partial S,
\end{equation}
and the angoricity \cite{angoricity},
\begin{equation}
A=\partial \Gamma/\partial S.
\end{equation}


In a recent series of papers \cite{compactivity,briscoe,max,meyer,yin}
the compactivity was used to describe frictional and frictionless hard
spheres in the volume ensemble. Here, we test the validity of the
statistical approach in the combined pressure-volume ensemble to
describe deformable, frictionless particles, such as emulsion systems
jammed under osmotic pressure near the jamming transition
\cite{forcedist,six}.


In general, if the density of states $g(\Gamma,\phi)$ in the space of
jammed configurations (defined as the probability of finding a jammed
state at a given $(\Gamma,\phi)$ at $A=\infty$) is known, then
calculations of macroscopic observables, like pressure $p$ and average
coordination number $Z$ as a function of $\phi$, can be performed by
the canonical ensemble average \cite{gama,canonicaljam} at a given
volume:
\begin{equation}
  \langle p(\alpha,\phi) \rangle_{\rm ens} =\frac {1}{\mathcal{Z}}
\int_{0}^{\infty}p\,\,g(\Gamma,\phi)\,\,
    e^{-\alpha \Gamma}\,\,{\rm d}\Gamma,
  \label{Eq_edP}
\end{equation}
and
\begin{equation}
  \langle Z(\alpha,\phi) \rangle _{\rm ens} =\frac {1}{\mathcal Z}
\int_{0}^{\infty}Z \,\,g(\Gamma,\phi)\,\, e^{-\alpha \Gamma}\,\,{\rm d}\Gamma,
\label{Eq_edZ}
\end{equation}
where the canonical partition function is
\begin{equation}
  \mathcal{Z}=\int_{0}^{\infty}g(\Gamma,\phi) e^{-\alpha \Gamma}{\rm
    d}\Gamma,
\end{equation}
and the density of states is normalized as
$\int_{0}^{\infty}g(\Gamma,\phi){\rm d}\Gamma=1$. The inverse
angoricity is defined as
\begin{equation}
\alpha \equiv 1/A= \partial S/\partial
\Gamma.
\end{equation}

At the jamming transition the system reaches isostatic equilibrium,
such that the stresses are exactly balanced in the resulting
configuration, and there exists a unique solution to the interparticle
force equations satisfying mechanical equilibrium. It is well
known that observables present power-law scaling \cite{jpoint,powerlaw,mode2}:

\begin{equation}
  \langle p \rangle_{\rm dyn} \sim (\phi-\phi_{c})^{a}\,,
  \label{Eq_mdP}
\end{equation}
\begin{equation}
 \langle Z \rangle_{\rm dyn}-Z_{c} \sim
  (\phi-\phi_{c})^{b},
  \label{Eq_mdZ}
\end{equation}
where $a=3/2$ and $b=1/2$ for Hertzian spheres and $Z_{c}=6$ is the
coordination number at the frictionless isostatic point (J-point)
\cite{clarification}.  The average $\langle \cdots \rangle_{\rm dyn}$
indicates that these quantities are obtained by averaging over
packings generated dynamically in either simulations or experiments as
opposed to the ensemble average over configurations $\langle \cdots
\rangle_{\rm ens}$ of Eqs. (\ref{Eq_edP})--(\ref{Eq_edZ}). Comparing
the ensemble calculations, Eqs. (\ref{Eq_edP})--(\ref{Eq_edZ}), with
the direct dynamical measurements, Eqs.
(\ref{Eq_mdP})--(\ref{Eq_mdZ}), provides a basic test of the ergodic
hypothesis for the statistical ensemble.

Our approach is the following: We first perform an exhaustive
enumeration of configurations to calculate $g(\Gamma,\phi)$ and obtain
$\langle p(\alpha,\phi) \rangle_{\rm ens}$ as a function of $\alpha$
for a given $\phi$ using Eq. (\ref{Eq_edP}).
Then, we obtain the angoricity by comparing the pressure in the
ensemble average with the one obtained following the dynamical
evolution with Molecular Dynamics (MD) simulations. By setting
$\langle p(\alpha,\phi) \rangle_{\rm ens} = \langle p \rangle_{\rm
  dyn}$,
we obtain the angoricity as a function of $\phi$.
By virtue of obtaining $\alpha(\phi)$, all the other observables can
be calculated in the ensemble formulation.  The ultimate test of
ergodicity is realized by comparing the remaining ensemble observables
with the corresponding direct dynamical measures.


\section{ Potential Energy Landscape approach: Ensemble calculations}
\label{pel-c}

\subsection{Features of the Potential Energy Landscape}
\label{pel}

An appealing approach for understanding out-of-equilibrium systems is
to study the properties of the system's ``potential energy landscape''
(PEL) \cite{Wales1}, described by the $3N$-coordinates of all
particles in the multi-dimensional configuration space, or landscape,
of the potential energy of the system ($N$ is the number of
particles).
Characterizing such potential energy landscapes has become an
important approach to study the behaviour of out-of-equilibrium
systems.  For example, this approach has provided important new
insights into the origin of the unusual properties of supercooled
liquids, such as the distinction between ``strong'' and ``fragile''
liquids \cite{Stillinger1}.

In frictionless granular matter, the potential energy is well-defined
and each jammed configuration corresponds to one local minimum in the
PEL.  For small systems ($N \lessapprox 14$), it is possible to find
all the minima with current computational power \cite{ohern}.  For
somewhat larger systems $N\approx 30$, it is possible to obtain a
representative ensemble, without exhaustively sampling all the
states. Based on these stationary points, we test the combined
volume-stress ensemble. The following work is only valid for
frictionless systems where the potential energy of interaction is well
defined. Frictional grains are path dependent due to Coulomb friction
between particles and therefore not amenable to a PEL study since
there is no well defined energy of interaction.

\begin{figure}
\centerline{ \resizebox{7.0cm}{!} {
    \includegraphics{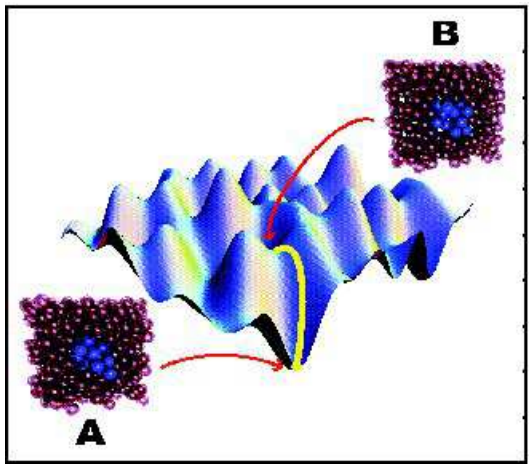}}}
\caption{A model two-dimensional potential energy surface.  The energy
  landscape is divided into basins of attraction, where the minima are
  the jammed states connected by pathways through saddle
  points. States A and B are typical packing configurations of 30
  particles (in blue) with their periodic boundary systems.}
\label{pel:pes}
\end{figure}

The formalism introduced by Goldstein \cite{Goldstein1} consists of
partitioning the potential energy surface into a set of basins as
illustrated in Fig. \ref{pel:pes}. The dynamics on the potential
energy surface can be separated into two types: the vibrational motion
inside each basin and the transitional motion between the local
minima. Stillinger and coworkers \cite{Weber1} developed the method of
inherent structure to characterize the PEL. In this method, a local
minimum in the PEL is located by following the steepest-descent
pathway from any point surrounding the minimum. The inherent structure
formalism simplifies the energy landscape into local minima and
ignores the vibrational motion around them. The dynamics between the
inherent structures is introduced with the transition states
identified with the saddle points in the PEL. The transition states
are stationary points like the local minima but they have at least one
maximum eigendirection.


\subsection{Finding Stationary States}

For the simplest system of $N$ structureless frictionless particles
possessing no internal orientational and vibrational degrees of
freedom,
the potential energy function of this N-body system is
$E(r_1,\ldots,r_N)$, where the vectors $r_i$ comprise position
coordinates. As mentioned above, the most interesting points of a
potential energy surface are the stationary points, where the gradient
vanishes. Here we explain how to locate these stationary points. The
algorithm follows well established methods in computational chemistry
\cite{Wales1}.  The procedure is analogous to finding the inherent
structures \cite{wales0} of glassy systems.  The algorithm employed,
LBFGS algorithm, is also similar to the conjugate gradient method
employed by O' Hern \cite{ohern,jpoint,manypackings}, differing in the
fact that it does not require the calculation of the Hessian matrix at
every time step.  We make the source code in C$++$ available at
http://www.jamlab.org and free to use together with all the packings
generated in this study.  The algorithm has been used in the short
version of this article \cite{wang-epl} and in a study of the PEL in
Lennard-Jones glasses to reconstruct a network of stationary states
and apply a percolation picture of the glass transition \cite{carmi}.

\subsection{General Method -- Newton-Raphson Method}

Consider the Taylor expansion of the potential energy, $E$, around a
general point in configuration space, $r$,
\begin{equation}
E(r+h) = E(r) + g^T h+\frac{1}{2}h^T H h + O(h^3),
\label{taylor}
\end{equation}
where $g$ is the gradient, $g_i =\partial_i E$, $H$ is the Hessian
matrix, $H_{ij} = \partial_i\partial_j E$, and $h$ is a
small step vector that gives the displacement away from $r$.

By Eq. (\ref{taylor}), the calculation of energy difference for a
given step $h$ from the initial point $r$ is complicated. By selecting
the eigenvectors of the Hessian matrix $e_{\alpha}$ as our local
coordinates, we can simplify the Taylor expansion of
Eq. (\ref{taylor}) as:
\begin{equation}
  \triangle E = E(r+h)-E(r) \approx \sum_\alpha
  (g_{\alpha}h_{\alpha}+\frac{\lambda_{\alpha}}{2}h_{\alpha}^2),
\label{changes}
\end{equation}
where $g=\sum_{\alpha} g_\alpha e_\alpha$, $h=\sum_{\alpha} h_\alpha
e_\alpha$, $H e_\alpha = \lambda_\alpha e_\alpha$, and
$\lambda_\alpha$ is the eigenvalue of the Hessian matrix for component
$\alpha$.

\begin{figure}
\centerline{ \resizebox{7.0cm}{!} {
\includegraphics{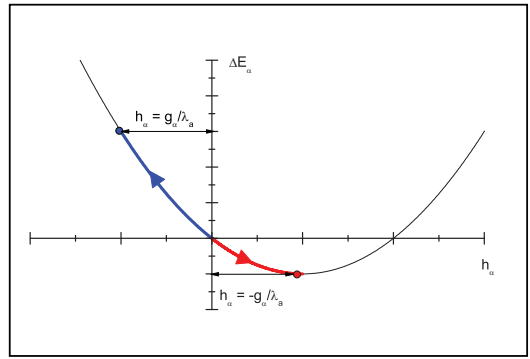}}}
\caption{A schematic energy change curve for one component with
  $\lambda_\alpha > 0$. We can select the downhill step as $h_\alpha =
  -\frac{g_\alpha}{\lambda_\alpha}$ to obtain a maximum energy change.
  The uphill step can not be too large since the Taylor expansion will
  not be accurate enough for the calculation.  Here, the uphill step
  is chosen as $h_\alpha = \frac{g_\alpha}{\lambda_\alpha}$.}
\label{pel:upanddown}
\end{figure}

From Eq. (\ref{changes}), it is easy to see that the total change of
energy could simply be the sum of the changes in each directions. This
may help us to raise the energy in some directions and reduce the
energy at others, and finally reach a stationary point. The length of
each step components can be selected as the maximum change of energy:
\begin{equation}
 h_\alpha = S_\alpha \frac{g_\alpha}{\lambda_\alpha},
\label{NRStep}
\end{equation}
as shown in Fig.\ref{pel:upanddown}. The sign $S_\alpha=\pm 1$ in this
formula depends on the choice of uphill or downhill direction. In
fact, for $\lambda_\alpha > 0$, it is possible to choose another step
for the uphill case, since $\triangle E_\alpha$ increases as $|
h_\alpha |$, but for large steps, the Taylor expansion Eq.
(\ref{taylor}) may breakdown. Therefore, it is important to control
the step length. For $\lambda_\alpha < 0$, we reach the opposite
conclusion.

The stationary points can be separated into local
minima and saddle points. Based on the eigenvalues of the Hessian
matrix for the stationary point, the local minima are ordered as:
\begin{equation}
0 \le \lambda_{1} \le \lambda_{2} \cdots \le \lambda_{3N},
\end{equation}


and for a saddle point of order $\alpha$:
\begin{equation}
  \lambda_{1} \le \cdots \le \lambda_{\alpha} \le 0 \le \lambda_{\alpha+1} \le \cdots \le \lambda_{3N}.
\end{equation}

Generally, this algorithm searches for the nearest stationary point on
the surface by following the opposite $(\lambda_{\alpha} \ge 0)$ and
along $(\lambda_{\alpha} \le 0)$ the various gradient directions.

\subsection{Finding local minima -- LBFGS algorithm}


It is much easier to locate local minima than saddle points
  because, for the first, we only need to search downhill in every
  direction.  At present one of the most efficient methods to search
  the local minima for large system is Nocedal's limited memory
  Broyden-Fletcher-Goldfarb-Shanno algorithm (LBFGS)
  \cite{Nocedal1,wales0}. The LBFGS algorithm constructs an
  approximate inverse Hessian matrix from the gradients (first
  derivatives) which are calculated from previous points. Since it is
  only necessary to calculate the gradients at each searching step,
  the LBFGS algorithm increases the computational speed of the
  algorithm enormously.

In the Newton-Raphson method discussed above, the Hessian matrix of
second derivatives is needed to be evaluated directly. Instead, the
Hessian matrix used in LBFGS method is approximated using updates
specified by gradient evaluations. The LBFGS algorithm code can be
obtained from http://www.netlib.org/opt/index.html.  Here we present a
brief explanation of the algorithm

From an initial random point $r_0$ and an approximate Hessian matrix
$H_0$ (in practice, $H_0$ can be initialized with $H_0=I$), the
following steps are repeated until $r$ converges to the local minimum.
\begin{itemize}
\item Obtain a direction $h_k$ by solving: $H_kh_k=-\nabla E(r_k)$.
\item Perform a line search to find an acceptable step size $\gamma_k$
  in the direction found in the first step, then update
  $r_{k+1}=r_k+\gamma_k h_k$.
\item Set $s_k=\alpha_kh_k$.
\item Set $y_k=\nabla E(r_{k+1})-\nabla E(r_k)$.
\item Set the new Hessian, $H_{k+1}=H_{k}+\frac{y_k
    y^T_k}{y^T_ks_k}-\frac{H_ks_k(H_ks_k)^T}{s^T_kH_ks_k}$.
\end{itemize}

\subsection{Finding saddles -- Eigenvector following method}

In the present study we do not make use of the saddle points. However,
other studies using network theory to represent the PEL necessitate the
links between minima through the saddle points \cite{carmi}. For
completeness, below we explain how to search for saddles.  A
particular powerful method for locating saddle points is the
eigenvector following method \cite{Wales1}.

The eigenvector-following method, developed by Cerjan, Miller and
  others \cite{Miller1,Wales1,Grigera1,Doye1,Wales2,Walsh1}, consists
  of locating a saddle point from a local minimum.  At each searching
  step towards a saddle point with $\alpha$ order, the directions are
  separated into two types: $\alpha$ uphill directions to maximization
  and $3N-\alpha$ downhill directions to minimization.

  We follow the implementation of the eigenvector-following method by
  Grigera \cite{Grigera1}. We give a general description: at each
  searching step, a step size $h$ is calculated by the diagonalized
  Hessian matrix \cite{Grigera1,Wales2,Walsh1}:
\begin{equation}
  h_\alpha = S_\alpha {2 g_\alpha \over |\lambda_\alpha| \left( 1 + \sqrt{1 + 4 g^2_\alpha
        / \lambda^2_\alpha } \right) },
\end{equation}
where $\lambda_\alpha$ are the eigenvalues of the Hessian matrix and
$g_\alpha$ are the components of the gradient in the diagonal base
($h_\alpha$ is set to 0 for the directions where
$\lambda_\alpha=0$). The sign $S_\alpha=\pm 1$ is chosen by the order
of the saddle point. For a saddle point of order $n$, the algorithm
will set $S_\alpha=-1$ for $1\le\alpha\le n$ and $S_\alpha=1$ for
$\alpha>n$.

When $g_\alpha\to 0$, the step size $h_\alpha$ converges to the
Newton-Raphson step as Eq. (\ref{NRStep}):
\begin{equation}
  h_\alpha = S_\alpha {g_\alpha \over \lambda_\alpha} + O(g_\alpha^2).
\end{equation}

\subsection{An Example}
\label{example}

We generate a two dimensional soft-ball system in circular boundary,
which contains 31 particles of equal radius, to illustrate the method
of finding stationary and saddle points in the PEL. The interaction
between particles (also the interaction between particles and
wall) follows the Hertzian law \cite{powerlaw}:
\begin{equation}
V(r_i,r_j) =\epsilon|r_i-r_j|^\frac{5}{2}
\end{equation}
Here, $\epsilon$ is the interaction strength between particles $i$ and
$j$.  The volume fraction is $\phi = 0.80$, which is closed to the
jamming transition of $2d$ hard disks.

\begin{figure}
\centerline{ \resizebox{5.0cm}{!} {
\includegraphics{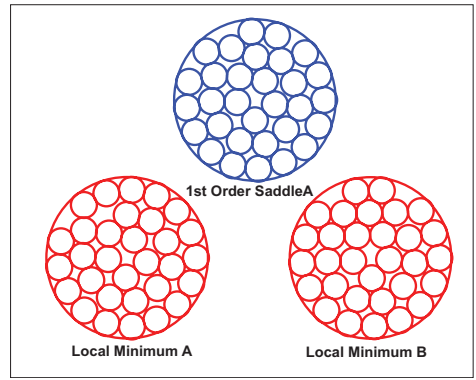}}}
\caption{A two dimensional 31 particle system in a circular
  boundary. Three different configurations in this system are
  generated with different algorithms. The LBFGS method is applied to
  locate minima A and B. For saddle C which connects A and B, the
  eigenvector following method is used.}
\label{pel:exp1}
\end{figure}

We first generate a random configuration, which is the initial point
of the search of the minima. With the LBFGS method, we search the
local minimum $A$ nearby this initial point. After the minimum $A$ is
obtained, we apply the eigenvector following method to walk from the
point $A$ on the potential energy surface to locate the transition
state $C$ (here the transition state is a first order
saddle). Finally, the minimum $B$ is located by applying LBFGS method
again. Figure \ref{pel:exp1} shows configurations of two local minima
(marked as red) and the transition state (marked as blue) between
them.

\begin{figure}
  \centerline{ \resizebox{7.0cm}{!} {
      \includegraphics{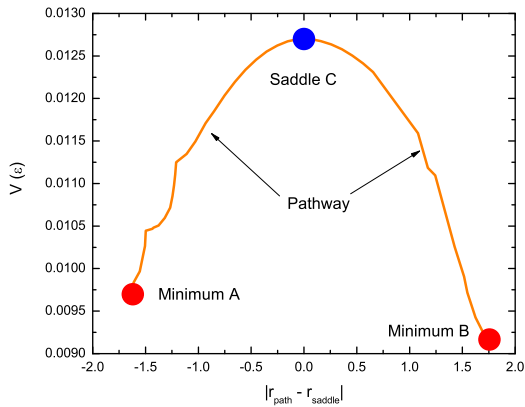}}}
  \caption{The pathway from minimum A to minimum B, passing by the
    saddle C, the x-coordinate is the distance from saddle C, the
    y-coordinate is the potential energy of the packing.}
\label{pel:exp2}
\end{figure}

The pathway from minimum A to minimum B, passing by transition state C, is
shown in Fig. \ref{pel:exp2}. The pathway distance is the Euclidean distance,

\begin{equation}
  d=\sqrt{(r'-r)\dot(r'-r)}
   =\sqrt{\sum_{i,\alpha} (r'_{i,\alpha} - r_{i,\alpha})^2 },
\end{equation}
where $i=1,2,3$, $\alpha=1 \cdots 3N$, $r'$ is the coordinate of configuration passing along the searching method and $r$ is the coordinate of saddle $C$.

The dynamics from minimum to minimum can be represented as a walk on a
network whose nodes correspond to the minima and where edges link
those minima which are directly connected by a transition state. The
work of Doye \cite{Doye2} provides an illustration of such a landscape
network for a LJ energy surface.  To characterize the topology of the
landscape network, Doye \cite{Doye2} study small Lennard-Jones
clusters to locate nearly all the minima and transition states on the
potential energy landscape. The inherent structure network of such a
system has a scale-free and small-world properties.  In a companion
study \cite{carmi} we repeated the main results as Doye studied. The
numbers of minima and transition states are expected to increase
roughly as $N_{min}\sim e^{\alpha N}$ and $N_{st}\sim Ne^{\alpha N}$
respectively, where $N$ is the number of atoms in the cluster.
Therefore, the largest network that we are able to consider is for a
14-atom cluster for which we have located 4158 minima and 90 738
transition states in agreement with the results of Doye.  In the next
Section we apply the above formalism to find the stationary states for
a 3d granular system of Hertz spheres in a periodic boundary.


\section{System Information. Hertzian system of spheres }
\label{hertz}

Next we calculate the density of jammed states $g(\Gamma,\phi)$ in the
framework of the PEL formulation for a system of Hertz spheres.  In
the case of frictionless jammed systems, the mechanically stable
configurations are defined as the local minima of the PEL
\cite{jpoint,manypackings}.

The systems used for both, ensemble generation and molecular dynamic
simulation, are the same. They are composed of 30 spherical particles
in a periodic boundary box. The particles have same radius $R = 5 \mu
m$ and interact via a Hertz normal repulsive force without friction.
The normal force interaction is defined as
\cite{powerlaw,hmlaw1,hmlaw2}:
\begin{equation}
F_n = \frac{2}{3}~ k_n R^{1/2} (\delta r) ^{\delta},
\end{equation}
where $\delta=3/2$ is the Hertz exponent, $\delta r= (1/2)[2 R -
|\vec{x}_1 -\vec{x}_2|]>0$ is the normal overlap between the spheres
and $k_n=4 G / (1-\nu)$ is defined in terms of the shear modulus $G$
and the Poisson's ratio $\nu$ of the material from which the grains
are made. We use typical values for glass: $G=29$ GPa and $\nu = 0.2$
and the density of the particles, $\rho=2 \times 10^{3}$ kg/m$^3$
\cite{powerlaw,hmlaw2}.  The interparticle potential energy is
\begin{equation}
  E=   \frac{2}{3}~ \frac{k_n}{\delta+1} R^{1/2} (\delta r)^{\delta+1}.
\end{equation}

The Hertz potential is chosen for its general applicability to
granular materials.  The results are expected to be independent of the
form of the potential. Below, we apply the LBFGS algorithms
\cite{Nocedal1,wales0} to find the local minima of the PES (zero-order
saddles).


\section{Ensemble Generation}
\label{dos}

In this section, we first explain the method to obtain geometrically
distinct minima in the PEL to calculate the density of states. Then we
show that the density of the states, $g(\Gamma,\phi)$, does not change
significantly after sufficient searching time for the configurations.

In principle, if all local minima corresponding to the mechanically
stable configurations of the PEL are obtained, the density of states
$g(\Gamma,\phi)$ can be calculated.  Such an exhaustive enumeration of
all the jammed states requires that $N$ not be too large due to
computational limits.  On the other hand, in order to obtain a precise
average pressure in the MD simulation, $\langle p \rangle_{\rm dyn}$,
$N$ cannot be too small such that boundary effects are
minimized. Considering these constraints, we choose a $30$ particle
system.


In order to enumerate  the jammed states at a given volume fraction
$\phi$, we start by generating initial unjammed packings (not
mechanically stable) performing a Monte Carlo (MC) simulation at a
high, fixed temperature. The MC part of the method applied to the
initial packings assumes a flat exploration of the whole PEL. Every MC
unjammed configuration is in the basin of attraction of a jammed state
which is defined as a local minimum in the PES with a positive
definite Hessian matrix, that is a zero-order saddle. In order to find
such a minimum, we apply the LBFGS algorithm provided by Nocedal and
Liu \cite{Nocedal1} explained above.  The PEL for each fixed $\phi$
likely includes millions of geometrically distinct minima by our
simulation results. Therefore, an exhaustive search of configurations
is computationally long; for a system of 30 particles it is impossible
to find all the configurations with the current available
computational power.  However, we notice that it not crucial to find
all the states, but rather a sufficiently accurate density of states.
Therefore, we check that the number of found configurations has
saturated after sufficient trials and that the density of states
$g(\Gamma,\phi)$ has converged to a final shape under a prescribed
approximation.

It is also important to determine if the local minima are
distinct. Usually, the eigenvalues of the Hessian matrix at each local
minimum can be used to distinguish these mechanically stable
packings. Here, we follow this idea to compare minima to filter the
symmetric packings. However, instead of calculating the eigenvalues of
each packing, which is time consuming, we calculate a function of
the distance between any two particles in the packing to improve
search efficiency (for the LBFGS algorithm, we do not need to
calculate the Hessian matrix). For each packing, we assign the
function $Q_{i}$ for each particle:
\begin{equation}
  Q_{i}=\sideset{}{}
  \sum_{1 \le j \le N,\;\ j\neq i} {\tan}^{2}(\frac {\pi r_{ij}^{2}}{3 L^{2}}), \label{Eq_tot}
\end{equation}
where $r_{ij}$ is the distance between particles $i$ and $j$, $L$ is
the system size and $N=30$. We list the $Q_{i}$ for each packing from
minimum to maximum $\lbrace Q_{i}\rbrace (1\le i \le N)$. Since $Q_{i}$
is a higher order nonlinear function, we can assume that two packings
are the same if they have the same list. The tolerance is defined as:
\begin{equation}
T =\sqrt {\frac {\sum_{\substack{1\le i \le N}}
(Q_{i}-Q^{'}_{i})^{2}}{N^{2}}},
\end{equation}
where $Q_{i}$ and $Q^{'}_{i}$ are the corresponding values from the
lists of two packings.

\begin{figure}
  \resizebox{7cm}{!}{\includegraphics{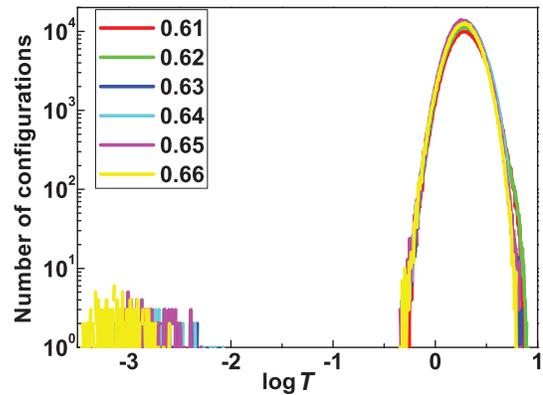}}
  \centering \caption{The distribution of the tolerance $T$ between
    any two packings at the given $\phi$. From the graph, the value of
    $T$ for which any two different packings are considered to be same
    is chosen to be $10^{-1}$, which is above the noise threshold and
    below the distribution of $T$.} \label{tolerance}
\end{figure}

Figure \ref{tolerance} shows the distributions of the tolerance $T$
for packings at different volume fractions. This figure suggests that
two packings can be considered the same if $T \le 10^{-1}$, which
defines the noise level.


\begin{figure}
  \resizebox{8cm}{!}{\includegraphics{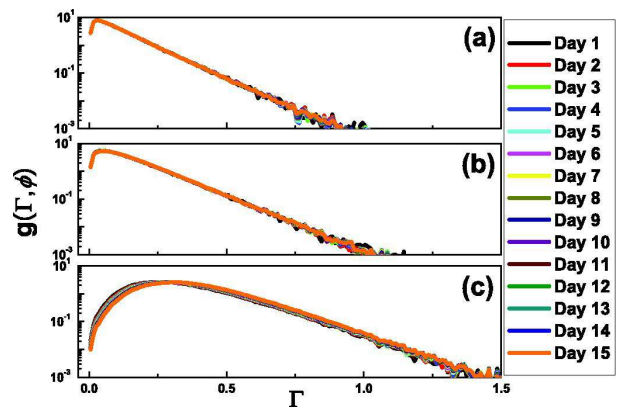}}
  \centering \caption{Log-log plot of the distribution of
    $g(\Gamma,\phi)$ for 15 searching days (a) at $\phi=0.609$, (b) at
    $\phi=0.614$, (c) at $\phi=0.625$. Different color in (a), (b),
    (c) corresponds to the different day. We find that after 15 days
    the distributions have converged.} \label{dos_phi}
\end{figure}

\begin{figure}
  \centerline{(a)
    \resizebox{6cm}{!}{\includegraphics{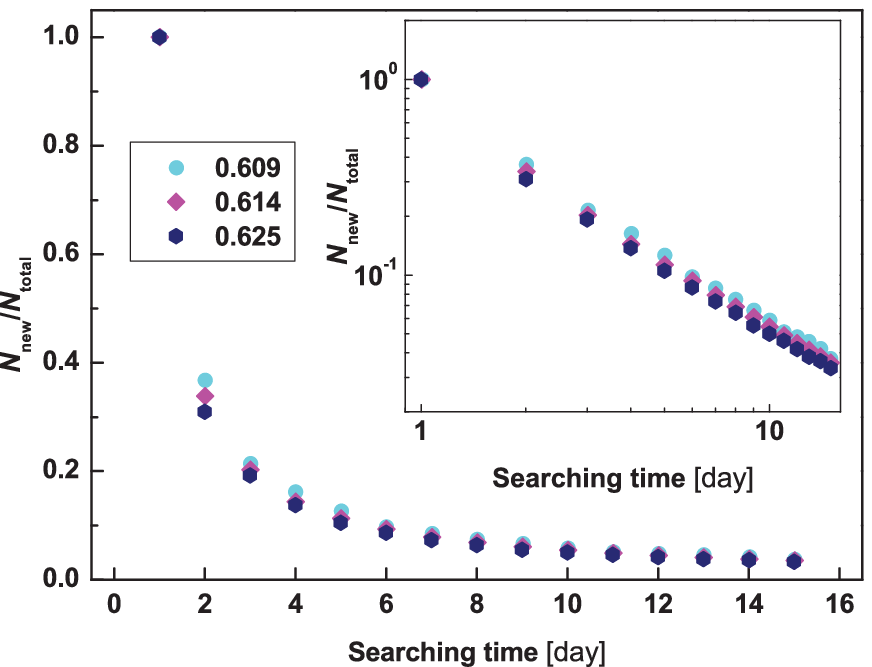}}}
  \centerline{(b) \resizebox{6cm}{!}{\includegraphics{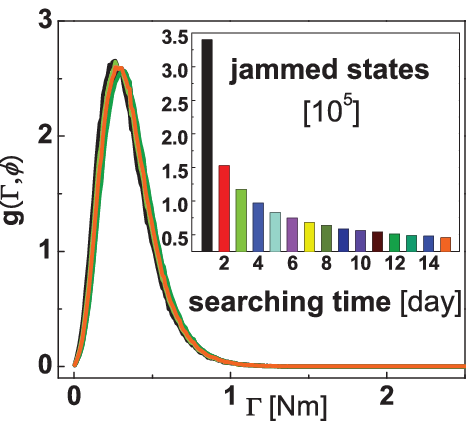}}}
  \centering \caption{(a) The probability to find new configurations
    as a function of searching time.  (b) Linear plot of the density
    of states as a function of searching time. Different colors
    indicate different days according to the inset. Inset shows the
    actual number of new configurations. } \label{number}
\end{figure}

\begin{figure}[h]
\centerline{\includegraphics[width=.8\linewidth]{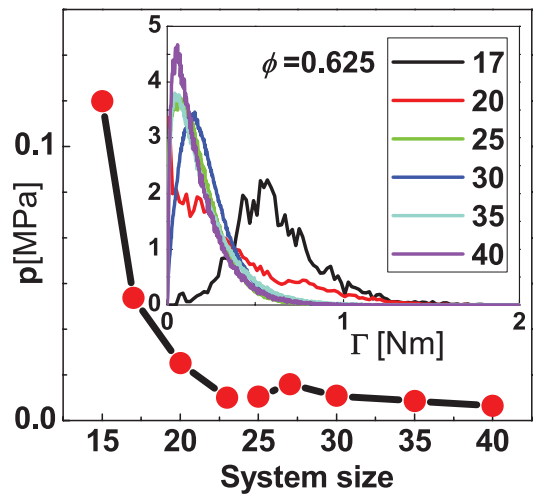}
}
\caption{Dependence of the results on the system size. The average
  value of $p$ converges as early as $N\sim 25$ particles.  The
  distribution $g(\Gamma,\phi)$ (inset) has not fully converged yet
  but its shape has converged after $N=35$ and the first moment does
  not change as indicated by the average $p$.  } \label{dos_figure_d}
\end{figure}

From Fig. \ref{dos_phi}, we see that after one week of searching,
$g(\Gamma,\phi)$ does not change significantly, since the initial
packings are generated by a completely random protocol.  We also
calculate the probability of finding new mechanically stable states
for different searching days, defined as $N_{\rm new}(i) / N_{\rm
  total}(i)$, where $N_{\rm new}(i)$ is the number of new
configurations found on the $i$-th day and $N_{\rm total}(i)$ is the
total number of configurations found in $i$ days. From
Fig. \ref{number}, we see that, after one week searching, the
probability of finding new configurations at different volume
fractions seems to have converged in the linear plot.  Figure
\ref{number}b shows a detail of the actual number of new
configurations found and $g(\Gamma,\phi)$ versus searching time in
days suggesting convergence. However, the log-log plot of the inset in
Fig. \ref{number}a indicates that the algorithm is still searching for
new configurations; the power-law relation in the inset suggesting a
neverending story.  However, the main question is whether the
observables have converged.  A further test of convergence is obtained
below in Fig. \ref{checkalpha} where the value of the inverse
angoricity is measured as a function of the searching time in days.
This plot suggests that enough ensemble packings have been obtained to
capture the features of $g(\Gamma,\phi)$ that give rise to the correct
observables.  We conclude that we have obtained an accurate enough
density of states for this particular system size.
Regarding system size dependence, the presented results are still $N$
dependent, although they started to converge for $N\sim 35$ and above,
Fig.~\ref{dos_figure_d}. More accurate calculations for large values
of $N$ remain computationally impossible, but in our treatment the
exact choice of $N$ is not as important as the consistency of the
results between ensemble and MD, for a given $N$ value.

\begin{figure}[h]
\centerline{\includegraphics[width=1.0\linewidth]{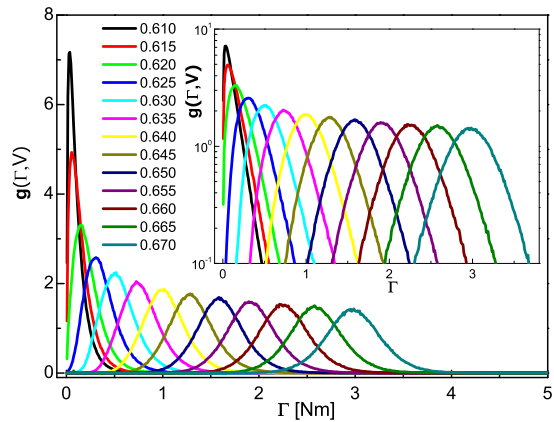}
}
\caption{The density of states $g(\Gamma,\phi)$ as a function of
  internal virial $\Gamma$ for different volume fraction, $\phi$,
  ranging from 0.610 to 0.670. The inset shows the logarithmic
  distribution of $g(\Gamma,\phi)$.  At low volume fraction ($\phi
  \lesssim 0.625$), the distributions are sharp and the tails of the
  distributions are exponential. At high volume fraction ($\phi
  \gtrsim 0.640$), the distributions are much broader and the tails
  are Gaussian.}
\label{dos_figure}
\end{figure}

Figure \ref{dos_figure} shows $g(\Gamma,\phi)$ versus $\Gamma$ for
different volume fractions.

\section{MD calculations}
\label{MD}

\begin{figure}
  \centerline{(a) \resizebox{6cm}{!}{\includegraphics{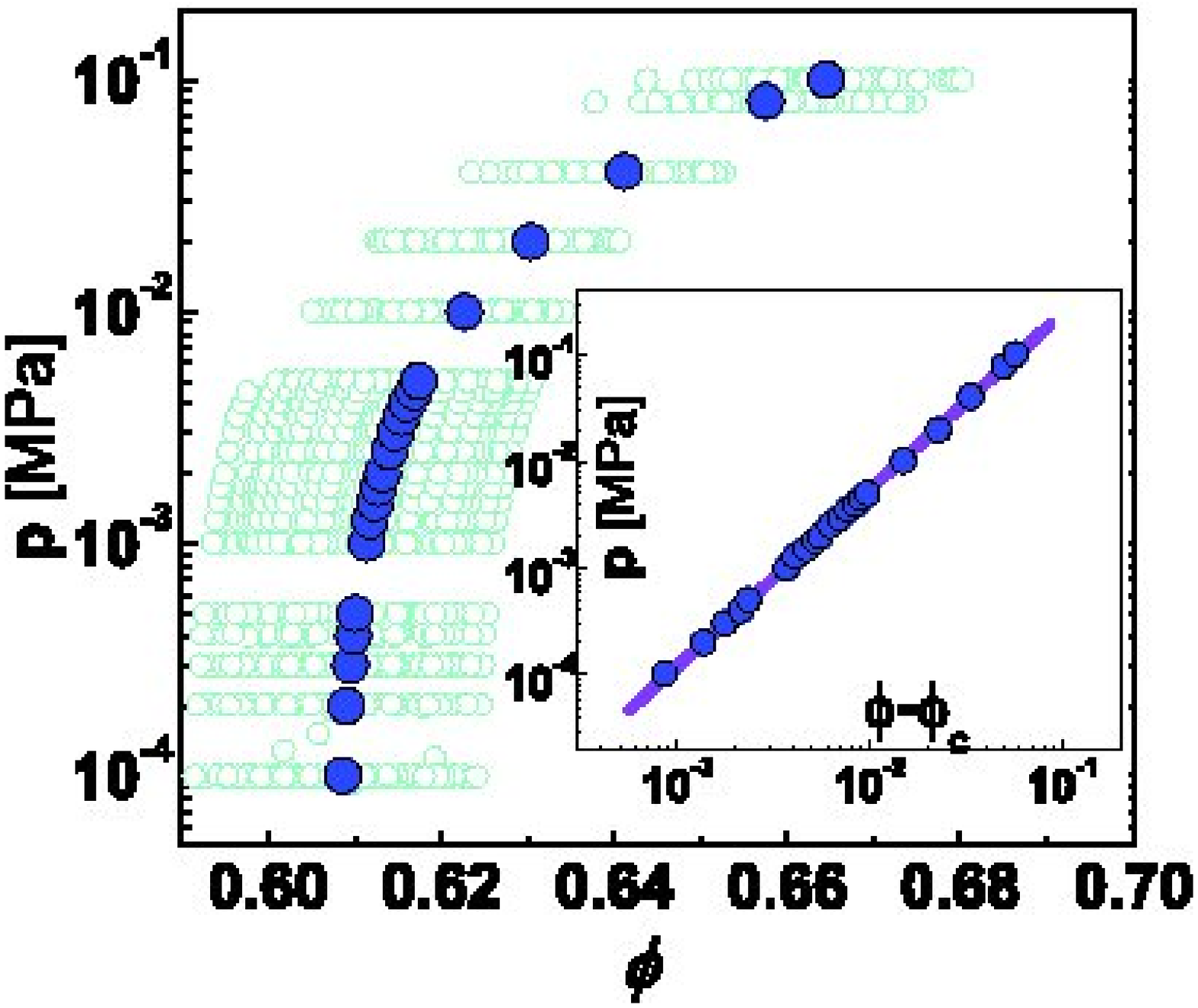}}}
  \centerline{(b) \resizebox{6cm}{!}{\includegraphics{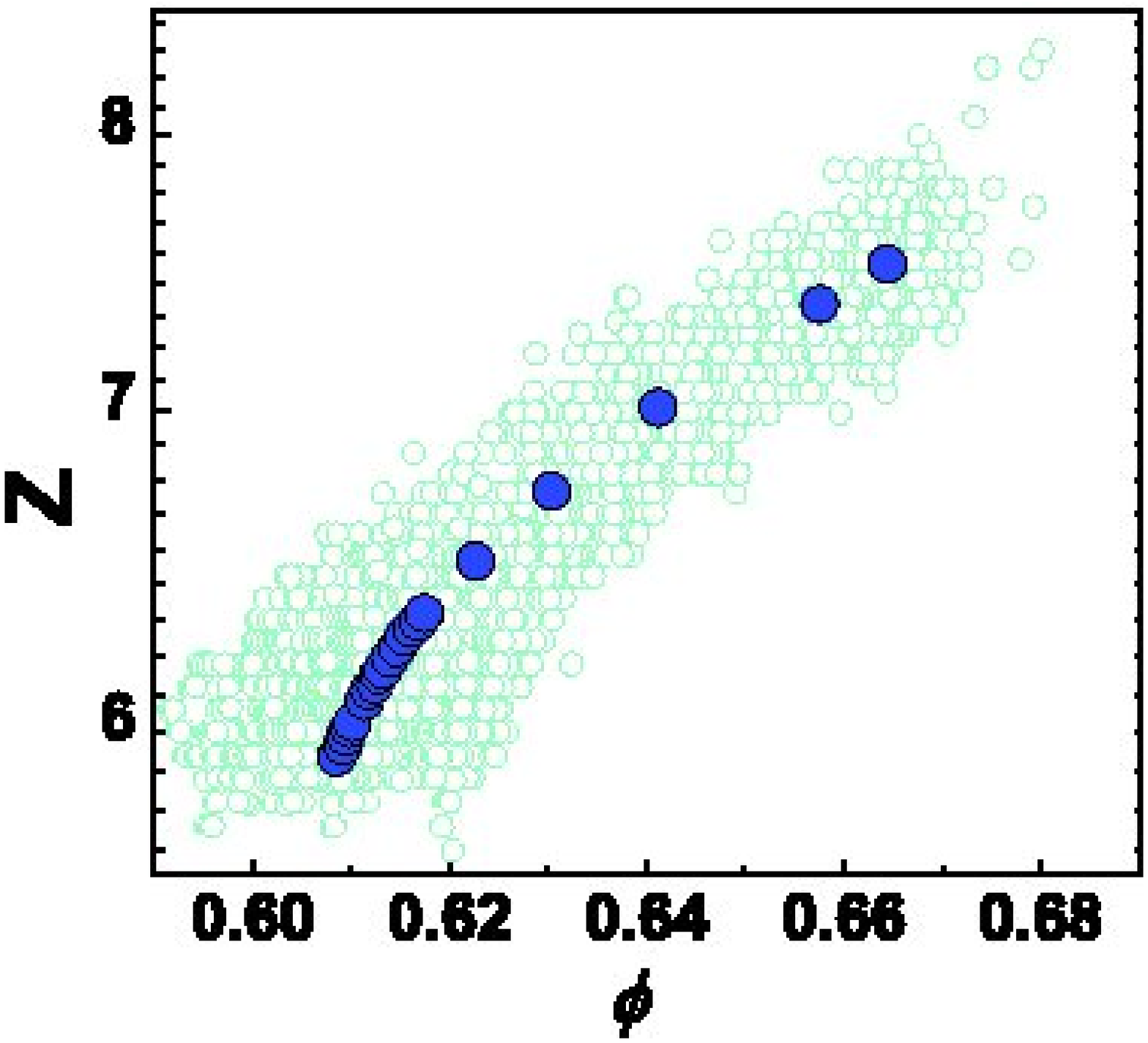}}}
  \centering \caption{The cyan $\bigcirc$ is (a) $\phi_{\rm dyn}$ and
    (b) $Z_{\rm dyn}$ for every single packing obtained with MD and
    the blue $\bigcirc$ is $\langle \phi \rangle_{\rm dyn}$ and
    $\langle Z \rangle_{\rm dyn}$ average over the single packings for
    the system which are shown in the text of the
    paper. } \label{all_md}
\end{figure}

In order to analyze numerical results, we perform MD simulations to
obtain $Z_{\rm dyn}$ and $\phi_{\rm dyn}$, which are
herein considered real dynamics. The algorithm is described in detail
in \cite{hmlaw2,compactivity,forcebalance2}.  Here, a general
description is given: A gas of non-interacting particles at an initial
volume fraction is generated in a periodically repeated cubic
box. Then, an extremely slow isotropic compression is applied to the
system. The compression rate is $\Gamma_0=5.9 t_0^{-1}$, where the
time is in units of $t_0=R\sqrt{\rho/G}$. After obtaining a state for
which the pressure $p$ is a slightly higher than the prefixed pressure
we choose, the compression is stopped and the system is allowed to
relax to mechanical equilibrium following Newton's equations. Then the
system is compressed and relaxed repeatedly until the system can be
mechanically stable at the predetermined pressure. To obtain the
statical average of $Z_{\rm dyn}$ and $\phi_{\rm dyn}$, we repeat the
simulation to get enough packing samples having statistically
independent random initial particle positions. Here, 250 independent
packings are obtained for each fixed pressure (see
Fig. \ref{all_md}). $\phi=\langle \phi \rangle_{\rm dyn}$ and $\langle
Z \rangle_{\rm dyn}$ are flat averages of these $250$ packings by
\begin{equation}
\langle \phi \rangle_{\rm dyn}=\frac {\sum_{1 \le i \le 250}
  \phi_{i}}{250},
\end{equation}
 and
\begin{equation}
\langle Z \rangle_{\rm dyn}=\frac {\sum_{1 \le
    i \le 250} Z_{i}}{250}.
\end{equation}

\begin{figure} [h]
  \resizebox{7cm}{!}{\includegraphics{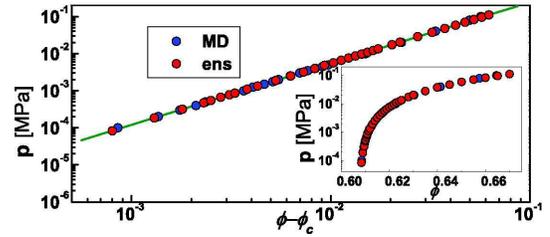}}
  \centering
  \caption{ Scaling of
    pressure.  The blue $\bigcirc$ shows the power-law relation for
    $\langle p \rangle_{\rm dyn}$ vs $\langle \phi \rangle_{\rm dyn}-
    \phi_{c} $ for the 30-particle system. Here, the pressure $\langle
    p \rangle_{\rm dyn}$ are average values obtained by 250
    independent MD simulations. The red $\bigcirc$ is the pressure
    used to obtain the inverse angoricity $\alpha$ predicted by
    Eq. (\ref{Eq_mdPvphi}). The relatively small system size results
    in large fluctuations of the observables. In order to predict a
    precise relation for the system ($N=30$), sufficient independent
    samples of the packings are generated to calculate the precise
    average for observables. We prepare 250 independent packings for
    each $\phi$ to get enough statistical samples to obtain $\langle p
    \rangle_{\rm dyn}$ and $\langle Z \rangle_{\rm dyn}$ by
    statistical average. The inset shows a semi-log plot.
  } \label{ango_a}
\end{figure}

 From previous studies, it has been observed the pressure $p$
 vanishes as power-law of $\phi$ when approaching the jamming
 transition as seen in Eq. (\ref{Eq_mdP}) \cite{jpoint,powerlaw}.
We obtain (Fig. \ref{ango_a})
\begin{equation}
  \langle p \rangle _{\rm dyn} = p_{0} \,\,
  ( \phi -\phi_{c})^{1.65} \,\,,
  \label{Eq_mdPvphi}
\end{equation}
where $ \phi_{c}=0.6077$ is the volume fraction corresponding to the
isostatic point J \cite{jpoint,powerlaw} following Eq. (\ref{Eq_mdZ})
and $p_{0}=10.8 {\rm MPa}$.  This critical value $\phi_c$ and the
exponent, $a = 1.65$, are slightly different from the values obtained
for larger systems ($a=\delta$) \cite{jpoint,powerlaw}. However, our
purpose is to use the same system in the dynamical calculation and the
exact enumeration for a proper comparison.

\section{Angoricity Calculation}

\label{angoricity}

Since we obtain $g(\Gamma,\phi)$ and $\langle p \rangle_{\rm dyn}$ for
each volume fraction $\phi$, we can calculate the inverse angoricity
$\alpha$ by Eq. (\ref{Eq_edP}). The pressure $\langle p(\alpha,\phi)
\rangle_{\rm ens}$ for a given $\phi$ is a function depending on
$\alpha$ as:

\begin{equation}
\langle p(\alpha,\phi)\rangle_{\rm ens} =\frac
{\int_{0}^{\infty}pg(\Gamma,\phi) e^{-\alpha \Gamma}{\rm
    d}\Gamma}{\int_{0}^{\infty}g(\Gamma,\phi) e^{-\alpha \Gamma}{\rm
    d}\Gamma} =\frac{\sum p e^{-\alpha \Gamma}}{\sum e^{-\alpha
    \Gamma}}.
  \label{Eq_sedP}
\end{equation}

\begin{figure} [h]
  \resizebox{7cm}{!}{\includegraphics{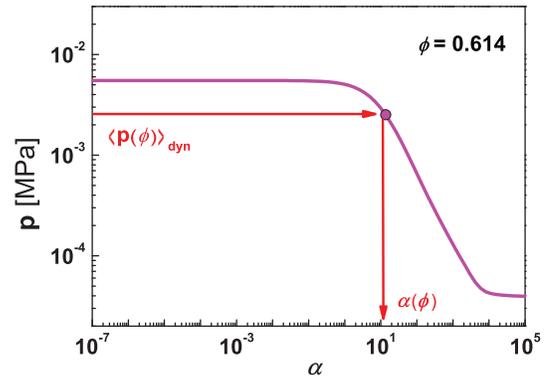}}
  \centering \caption{The numerical integration of Eq. (\ref{Eq_sedP})
    for $\phi=0.614$ is shown as the pink curve. We input the $\langle
    p \rangle_{\rm dyn}$ (pink $\bigcirc$ in the plot) and obtain the
    corresponding inverse angoricity $\alpha$.} \label{alpha_phi}
\end{figure}

\begin{figure} [h]
  \resizebox{7cm}{!}{\includegraphics{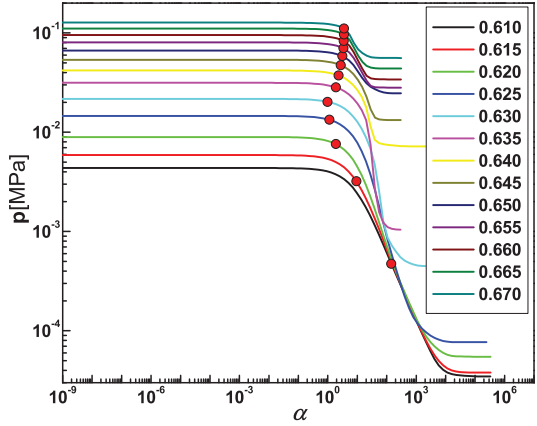}}
  \centering \caption{Calculation of $\alpha$ for several volume
    fractions $\phi$ as explained in detail in
    Fig. \ref{alpha_phi}} \label{p_vs_alpha}
\end{figure}

Figure \ref{alpha_phi} shows the result of the numerical integration
of Eq. (\ref{Eq_sedP}) for a particular $\phi=0.614$ as a function of
$\alpha$ using the numerically obtained $g(\Gamma,\phi)$ from
Fig. \ref{dos_figure}. To obtain the value of $\alpha$ for this
$\phi$, we input the corresponding measure of the pressure obtained
dynamically $\langle p(\phi) \rangle_{\rm dyn}$ and obtain the value
of $\alpha$ as schematically depicted in Fig. \ref{alpha_phi}.  The
same procedure is followed for every $\phi$ (see
Fig. \ref{p_vs_alpha}) and the dependence $\alpha(\phi)$ is
obtained.

\begin{figure} [h]
  \resizebox{7cm}{!}{\includegraphics{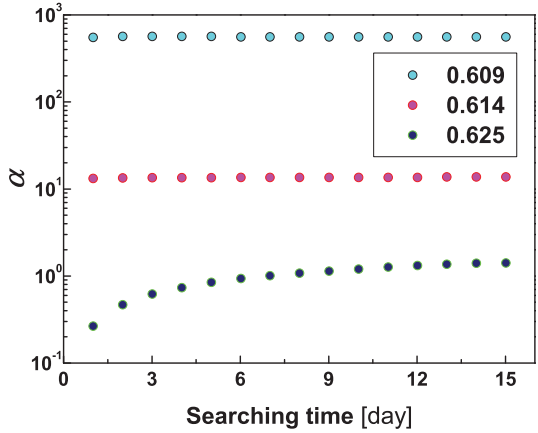}}
  \centering \caption{Calculation of inverse angoricity $\alpha$ as a
    function of searching time.} \label{checkalpha}
\end{figure}

We also check the inverse angoricity $\alpha(\phi)$ using
$g(\Gamma,\phi)$ for different searching days.
to ensure the accuracy and convergence to the proper value. From
Fig. \ref{checkalpha}, we can see that, after 10 days searching,
$\alpha(\phi)$ is stable due to the fact that the density of state,
$g(\Gamma,\phi)$, does not change significantly.

\begin{figure}
 \centerline{ (a) \resizebox{6cm}{!}{\includegraphics{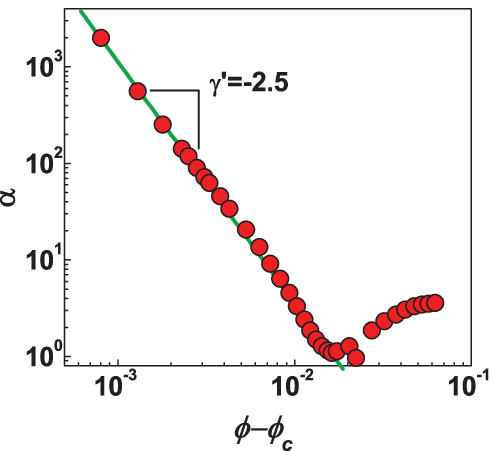}}}
\centerline{  (b) \resizebox{6cm}{!}{\includegraphics{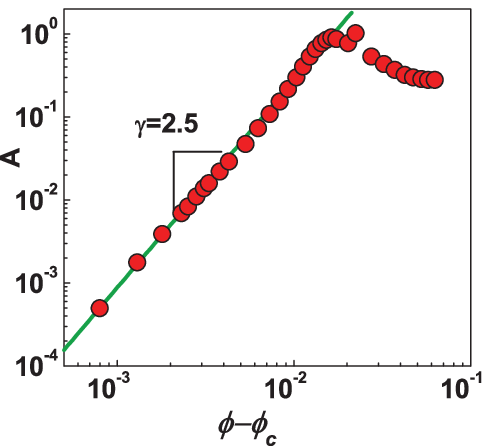}}}
\caption{(a) Inverse angoricity $\alpha$ as a function of
  $\phi$-$\phi_{c}$. We find a power-law relation for system's volume
  fraction $\phi$ near $\phi_{c}$. The solid line has a slope of
  -2.5. (b) The angoricity $A(=1/\alpha)$ vs $\phi$-$\phi_{c}$. To
  find $A$ accurately for system's volume fraction $\phi$ much larger
  than $\phi_{c}$, becomes difficult due to the large fluctuations and
  finite size effects. In principle, we expect that the plateau of $A$
  for large volume fraction $\phi$ might be related to the finite size
  of the sample. Indeed it is very difficult to estimate $\alpha$
  since it falls in the plateau in Fig. \ref{p_vs_alpha}.
} \label{ango_b}
\end{figure}

For each $\phi$ we use $g(\Gamma,\phi)$ to calculate $\langle p
(\alpha)\rangle_{\rm ens}$ by Eq. (\ref{Eq_edP}). Then, we obtain
$\alpha(\phi)$ by setting $\langle p(\alpha,\phi) \rangle_{\rm ens} =
\langle p \rangle_{\rm dyn}$ for every $\phi$. The resulting equation
of state $\alpha(\phi)$ is plotted in Fig. \ref{ango_b} and shows that
the angoricity follows a power-law, near $\phi_c$, of the form:
\begin{equation}
A \propto
(\phi-\phi_{c})^{\gamma},
\label{inverse}
\end{equation}
with $\gamma = 2.5$. The result is consistent with $\gamma = \delta +
1.0$, suggesting that $A \propto \Gamma \propto F_{n} r$.  For volume
fraction much larger than $\phi_c$, the system's input pressure
$\langle p(\phi) \rangle_{\rm dyn}$ reaches the plateau at low
$\alpha$ of the function $\langle p(\alpha,\phi)\rangle_{\rm ens}$
(see Fig. \ref{p_vs_alpha}) and the corresponding $\alpha(\phi)$
becomes much smaller (the angoricity $A(\phi)$ becomes much larger),
leading to large errors in the value of $A$ as $\phi$ becomes
large. This might explain the plateau found in $A$ when $(\phi -
\phi_c) > 2 \times 10^{-2}$ as shown in Fig. \ref{ango_b}.

Angoricity is a measure of the number of ways the stress can be
distributed in a given volume. Since the stresses have a unique
solution for a given configuration at the isostatic point, $\phi_c$,
the corresponding angoricity vanishes.
At higher pressure, the system is determined by multiple degrees of
freedom satisfying mechanical equilibrium, leading to a higher stress
temperature, $A$.
The angoricity can also be viewed as a scale of stability for the
system at different volume fractions. Systems jammed at larger volume
fractions require higher angoricity (higher driving force) to rearrange.

\section{Test of ergodicity}
\label{ergodicity}

In principle, using the inverse angoricity, $\alpha$, from
Eq. (\ref{inverse}) we can calculate any macroscopic statistical
observable $\langle B \rangle_{\rm ens}$ at a given volume by
performing the ensemble average \cite{canonicaljam}:
\begin{equation}
  \langle B(\phi) \rangle _{\rm ens} =\frac{1} {{\mathcal{Z}}}
  \int_{0}^{\infty}B \,\, g(\Gamma,\phi) \,\, e^{-\alpha \Gamma} \,\,{\rm d}\Gamma.
  \label{Eq_edaB}
\end{equation}
We test the ergodic hypothesis in the Edwards's ensemble
by comparing Eq. (\ref{Eq_edaB}) with
the corresponding value obtained with MD simulations averaged over
($250$) sample packings, $B_i$, generated dynamically:
\begin{equation}
  \langle B(\phi) \rangle _{\rm dyn} =\frac {1}{250}
  \sum_{i=1}^{250} B_{i}.
\label{Eq_mdaB}
\end{equation}

The comparison is realized by measuring the average coordination
number, $\langle Z \rangle$, the average force and the distribution of
interparticle forces. We calculate $\langle Z \rangle_{\rm ens}$ by
Eq. (\ref{Eq_edZ})
and $\langle Z \rangle_{\rm dyn}$ as in Eq. (\ref{Eq_mdaB}).
Using $\alpha(\phi)$ for each volume fraction, we calculate $\langle Z \rangle_{\rm ens}$ by:
\begin{equation}
  \langle Z(\phi) \rangle_{\rm ens} =\frac {\int_{0}^{\infty}Zg(\Gamma,\phi)
    e^{-\alpha \Gamma}{\rm d}\Gamma}{\int_{0}^{\infty}g(\Gamma,\phi)
    e^{-\alpha \Gamma}{\rm d}\Gamma}
  =\frac{\sum Z e^{-\alpha \Gamma}}{\sum e^{-\alpha \Gamma}}.
  \label{Eq_sedZ}
\end{equation}

\begin{figure} [h]
  \centerline{(a) \resizebox{6cm}{!}{\includegraphics{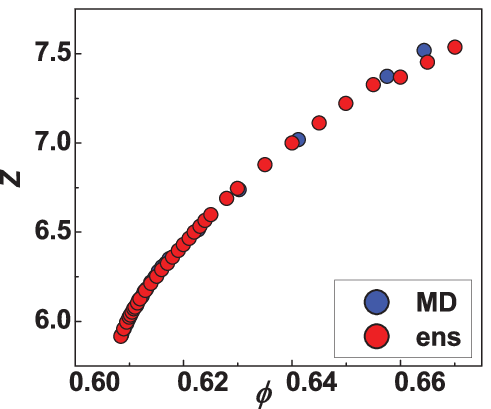}}}
  \centerline{(b) \resizebox{6cm}{!}{\includegraphics{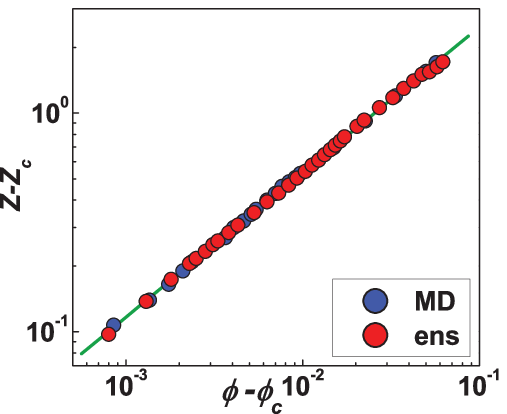}}}
  \centering \caption{ {\bf Test of ergodicity.}  (a) The blue
    $\bigcirc$ is the average coordination number $\langle Z \rangle
    _{\rm dyn}$ obtained by 250 independent MD simulations. The red
    $\bigcirc$ is the coordination number $\langle Z \rangle _{\rm
      ens}$ calculated by the ensemble for different volume
    fractions. Agreement between both measures supports the concept of
    ergodicity in the system.  (b)The same as (a) but in a log-log
    plot. The blue $\bigcirc$ shows the power-law relations for
    $\langle Z \rangle_{\rm dyn}$-$Z_{c}$ vs $\langle \phi
    \rangle_{\rm dyn}$ -$\phi_{c}$ for 30-particle system with
    $\phi_{c}=0.6077$ and $Z_{c}=5.82$.  } \label{predict_a}
\end{figure}

\begin{figure} [h]
\centerline{(a)  \resizebox{6cm}{!}{\includegraphics{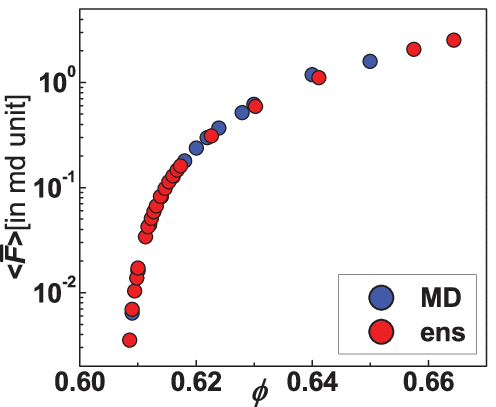}}}
\centerline{(b)  \resizebox{6cm}{!}{\includegraphics{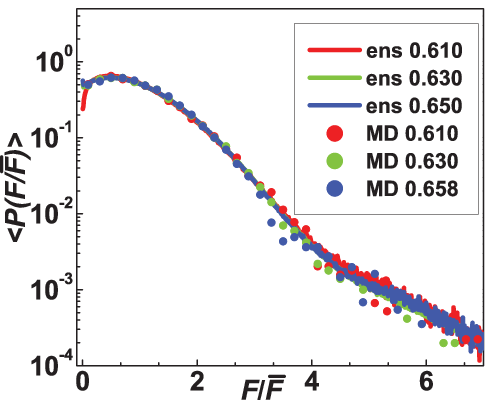}}}
  \centering \caption{
  {\bf Test of ergodicity.}  (a) Comparison of $\langle \overline F
  \rangle_{\rm dyn}$ and $\langle \overline F\rangle_{\rm ens}$ for
  different volume fractions.  (b) The comparison of selected
  distribution of force $ P _{\rm dyn}(F/\overline F )$ and $ P_{\rm
    ens}(F/\overline F ) $ for different volume fractions.
} \label{predict_b}
\end{figure}

The average force $\langle\overline F \rangle_{\rm ens}$ is given by:

\begin{equation}
  \langle\overline F(\phi) \rangle_{\rm ens} =\frac {\int_{0}^{\infty} \overline F g(\Gamma,\phi)
    e^{-\alpha \Gamma}{\rm d}\Gamma}{\int_{0}^{\infty}g(\Gamma,\phi)
    e^{-\alpha \Gamma}{\rm d}\Gamma}
  =\frac{\sum \overline F e^{-\alpha \Gamma}}{\sum e^{-\alpha \Gamma}},
  \label{Eq_sedF}
\end{equation}
where $\overline F$ is the average force for each ensemble
packing. Finally, the force distribution $P_{\rm ens}(F/\overline F)$
is given by:
\begin{equation}
   P _{\rm ens}(F/\overline F) =\frac {\int_{0}^{\infty}P(F/\overline F )g(\Gamma,\phi) e^{-\alpha \Gamma}{\rm d}\Gamma}{\int_{0}^{\infty}g(\Gamma,\phi)
    e^{-\alpha \Gamma}{\rm d}\Gamma}
    =\frac{\sum P(F/\overline F ) e^{-\alpha \Gamma}}{\sum e^{-\alpha \Gamma}}.
    \label{Eq_sedPro}
\end{equation}
Equations (\ref{Eq_sedZ})--(\ref{Eq_sedPro}) are then compared with
the dynamical measures for a test of ergodicity in
Figs. \ref{predict_a} and \ref{predict_b}.

Figure \ref{predict_a}a and \ref{predict_a}b show that the two independent
estimations of the coordination number agree very well: $\langle Z
\rangle _{\rm ens}=\langle Z \rangle _{\rm dyn}$.  The average
inter-particle force $\overline F $ for a jammed packing is
proportional to the pressure of the packing. We calculate
$\langle\overline F \rangle_{\rm ens}$ and
$\langle\overline F \rangle_{\rm dyn}$ and find that they coincide
very closely (see Fig. \ref{predict_b}a).  The full distribution of
inter-particle forces for jammed systems is also an important
observable which has been extensively studied in previous works
\cite{jpoint,forcedist1,forcedist2}. The force distribution is
calculated in the ensemble $P_{\rm ens}(F/\overline F)$ by averaging
the force distribution for every configuration in the PES.
Figure \ref{predict_b}b shows the distribution functions. The peak of
the distribution shown in Fig. \ref{predict_b}b indicates that the
systems are jammed \cite{jpoint,forcedist1,forcedist2}. Besides the
exact shape of the distribution, the similarity between the ensemble
and the dynamical calculations shown in Fig. \ref{predict_b}b is
significant.  The study of $\langle Z \rangle$, $\langle\overline F
\rangle$ and $P(F/\overline F)$ reveals that the statistical ensemble
can predict the macroscopic observables obtained in MD. We conclude
that the idea of ``thermalization'' at an angoricity is able to
describe the jamming system very well.

The MD simulations performed so far are at a predetermined pressure
$p$.  For this case there is no difference between the force
distribution $P(F/\overline F )$ and $P(F/\langle\overline F \rangle)$
\cite{jpoint}.  On the other hand, a MD simulation at a given fixed
volume fraction $\phi$, gives rise to different distributions.  For
each system with fixed $\phi$, the packings can have various
pressure. This suggests that the force distribution for each packing
scaled by the average force over all packings, $P(F/\langle\overline F
\rangle)$, should be different from the force distribution scaled by
the average force of that particular packing $P(F/\overline F )$
\cite{ohern}.  We now proceed to investigate a constant volume MD, vMD
simulation.

\begin{figure} [h]
  \resizebox{9cm}{!}{\includegraphics{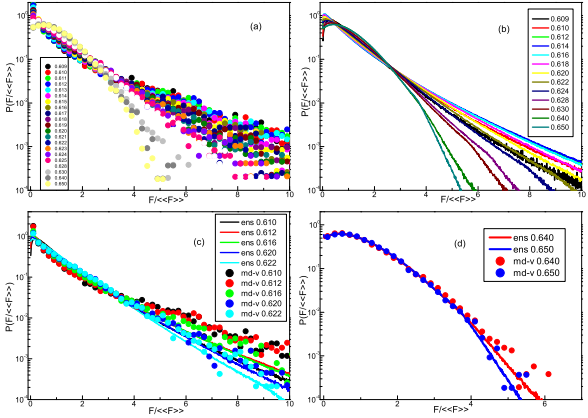}}
  \centering \caption{(a) The distribution of force $P_{\rm
      vMD}(F/\langle \overline F \rangle_{\rm vMD})$. (b) The
    distribution of force $P_{\rm ens}(F/\langle \overline F
    \rangle_{\rm ens})$. (c) and (d) The comparison of selected
    $P(F/\langle \overline F \rangle)$ between vMD and ensemble
    predicted by angoricity.} \label{pofff}
\end{figure}

\begin{figure} [h]
  \resizebox{7cm}{!}{\includegraphics{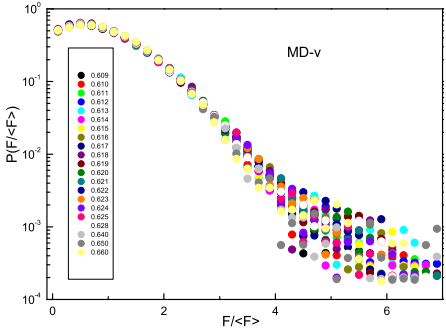}}
  \centering \caption{The distribution of forces, $P(F/\langle F
    \rangle)_{\rm vMD}$ } \label{poffv}
\end{figure}

The force distribution for vMD ensemble, $P_{\rm
  dyn}(F/\langle\overline F \rangle_{\rm dyn})$ is shown in
Fig. \ref{pofff}a. From Fig. \ref{pofff}a, we find that the force
distribution $P_{\rm dyn}(F/\langle\overline F \rangle_{\rm dyn})$ as
a function of different volume fraction $\phi$ no longer collapse. At
$\phi$ close to $\phi_{c}$, the average system force $\overline F$ for
each packing changes dramatically. While at $\phi$ is much above
$\phi_{c}$, the fluctuations of the average system force $\overline F
$ decrease, then the force distribution $P_{\rm
  dyn}(F/\langle\overline F \rangle_{\rm dyn})$ changes continuously.

We can also calculate the force distribution $P_{\rm
  ens}(F/\langle\overline F \rangle_{\rm ens})$ in the ensemble average:
\begin{equation}
  P _{\rm ens} (F/\langle\overline F \rangle_{\rm ens}=\frac {\int_{0}^{\infty}P(F/\langle\overline F \rangle_{\rm ens})g(\Gamma,\phi) e^{-\alpha \Gamma}{\rm d}\Gamma}
  {\int_{0}^{\infty}g(\Gamma,\phi) e^{-\alpha \Gamma}{\rm d}\Gamma}, \label{Eq_edaaF}
\end{equation}
where $\langle\overline F \rangle_{\rm ens}$ is the overall average
$\overline F $ of the ensemble.

From Fig. \ref{pofff}b, we find the same tendency as obtained in MD
simulation.  Furthermore, we check the distribution of force
$P(F/\langle F \rangle)$ for our vMD system (see Fig. \ref{poffv}).
We see that $P(F/\langle F \rangle)$ for different volume fraction
$\phi$ collapses very well similarly to those obtained from the
predetermined pressure system. This result suggests that $P(F/\langle
F \rangle)$ is a global quantity that can be used to verify if the
system is jammed or not \cite{ohern}.

\begin{figure} [h]
  \resizebox{7cm}{!}{\includegraphics{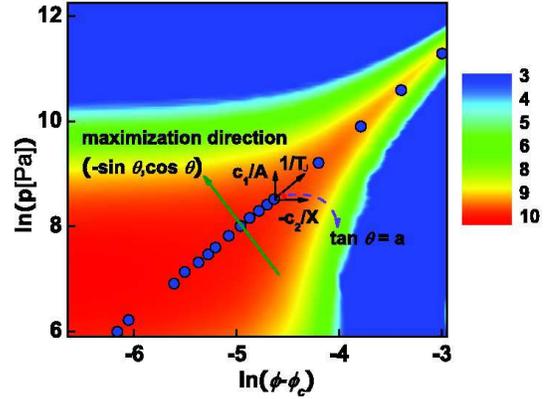}}
  \centering \caption{ Microcanonical calculations.  The
  entropy surface $S(\ln(\phi-\phi_{c}),\ln p)$. The color bar
  indicates the value of the entropy. The superimposed blue $\bigcirc$
  is $\langle p(\phi) \rangle_{\rm dyn}$ from MD calculations as in
  Fig. \ref{ango_a}. The olive arrow line indicates the maximization
  direction of the entropy $(-\sin \theta,\cos \theta)$. Following
  this direction, the entropy is maximum at the point $(\ln(\langle
  \phi \rangle _{\rm dyn}-\phi_{c}),\ln \langle p \rangle _{\rm
    dyn})$, corroborating the maximum entropy principle.
} \label{entropydis}
\end{figure}

\section{Thermodynamic analysis of the jamming transition}
\label{thermo}

So far we have considered how the angoricity determines the pressure
fluctuations in a jammed packing at a fixed $\phi$. The role of the
compactivity in the jamming transition can be analyzed in terms of the
entropy which is easily calculated in the microcanonical ensemble from
the density of states. Figure \ref{entropydis} shows the entropy of the
system as a function of  $(p,\phi)$ in phase space:
\begin{equation}
S=\ln(\Omega(p,\phi)).
\end{equation}
Here $\Omega$ is the number of states which is the unnormalized
version of $g(\Gamma,\phi)$.  It is important to note that
Fig. \ref{entropydis} shows the non-equilibrium entropy, in the
Edwards sense.  At the Edwards equilibrium, the entropy is maximum
respect to changes in $\phi$ and $\Gamma$. We will now see how the
jammed system verifies the principle of maximum entropy.

We analyze the entropy surface $S(\ln(\phi-\phi_{c}),\ln p)$ plotted
versus $(\ln(\phi-\phi_{c}),\ln p)$ in Fig.  \ref{entropydis}. When we
plot superimposed the MD-obtained curve $\langle p(\phi) \rangle_{\rm
  dyn}$ we see that the MD values pass along the maximum of the
entropy surface constrained by the coupling between $p$ and $\phi$,
Eq.  (\ref{Eq_mdZ}) (such a curve is superimposed to the entropy
surface in Fig. \ref{entropydis}). Due to the coupling through the
contact force law, the maximization of entropy is not on $p$ or $\phi$
alone but on a combination of both. The entropy $S$ reaches a maximum
at the point $S(\ln(\langle \phi \rangle _{\rm dyn}-\phi_{c}),\ln
\langle p \rangle _{\rm dyn})$ when we move along the direction
perpendicular to the jamming curve $\langle p (\phi) \rangle_{\rm
  dyn}$ (see the maximization direction in Fig.
\ref{entropydis}). This is a direct verification of the second-law of
thermodynamics: the dynamical measures maximize the entropy of the
system.

We can use this result to obtain a relation between angoricity and
compactivity and show how a new ``jamming temperature'' $T_{\rm J}$
and the corresponding jamming ``heat'' capacity $C_{\rm J}$ can
describe the jamming transition.

From the power-law relation $p=\Gamma/V \propto(\phi-\phi_{c})^{a}$,
we have:
\begin{equation}
 \ln p = \ln p_0 + a\ln(\phi-\phi_c), \label{Eq_powerlog0}
\end{equation}
where $p_0$ is the constant depending on the system.

\begin{figure} [h]
  \resizebox{8cm}{!}{\includegraphics{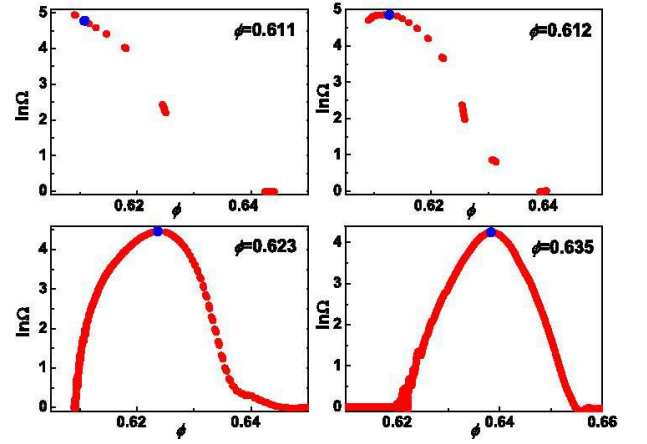}}
  \centering \caption{The non-equilibrium entropy $S(\ln p,
    \ln(\phi-\phi_{c}))$ along the direction $(-\sin \theta,\cos
    \theta)$ for different jamming ensemble points. The blue
    $\bigcirc$ represents the entropy of the jammed system obtained
    from MD. We see that closely follows the maximum of $S$ for all
    the volume fractions except very close to the jamming point where
    the blue point does not coincide with the maximum of $S$. It
    remains to be studied if this deviation is a finite size effect,
    or it could be due to a different value of $\phi_c$ between
    simulations and microcanonical ensemble.}
   \label{distentropy}
\end{figure}

Figure \ref{entropydis} indicates that the jammed system always remain
at the positions of maximal entropy,
\begin{equation} \delta S=0, \end{equation} in the direction ($-\sin
\theta$,$\cos \theta$), perpendicular to the jamming power-law curve
and the slope
\begin{equation}
\tan \theta = a.
\end{equation}
 In order to further analyze this result, we plot the entropy
distribution along the direction ($-\sin \theta$,$\cos \theta$) in
Fig.  \ref{distentropy}. We see that the entropy of the corresponding
jammed states remains at the peak of the distributions along ($-\sin
\theta$,$\cos \theta$).  This is clear when we plot the value of
$(p,\phi)$ from MD simulations in the plot of $S$ in
Fig. \ref{distentropy}, blue dot. Except for volumes very close to
jamming, the MD coincides with the maximum of $S$ when taken along
($-\sin \theta$,$\cos \theta$).  We notice that the maximization is
quite accurate for large volume fractions. For $\phi$ close to jamming
deviations are seen. We cannot rule out that these deviations are
finite size effects.  The deviations for small $\phi$
(Fig. \ref{distentropy}) remains to be studied. They could be due to
finite size effects or due to the fact that the value of $\phi_c$ is
different for the MD results and the microcanonical ensemble $S$ due
to the small size of the system.  In general, this plot verifies the
maximum entropy principle in this particular direction.  An analogous
plot where the entropy is shown as a function of $\phi$ but along the
horizontal direction (or along the vertical direction, $\Gamma$) shows
that the MD entropy is not maximal along these two directions.

\begin{figure} [h]
  \resizebox{7cm}{!}{\includegraphics{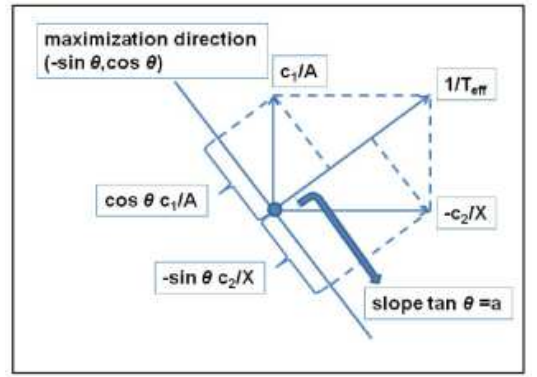}}
  \caption{The representation of the maximization analysis $\delta
    S=0$ along the direction $(-\sin \theta, \cos \theta)$ for one
    point in the jamming power-law curve. Here $c_1=\Gamma$ and
    $c_2=(\phi-\phi_{c})(N V_{g}/\phi^{2})$.}
   \label{slope}
\end{figure}

Thus, the maximization of entropy is not on $\Gamma$ or $V$ alone, but
on a combination of both.  This means that the entropy $S(\ln(\langle
\phi \rangle _{\rm dyn}-\phi_{c}),\ln \langle p \rangle _{\rm dyn})$
is maximum along the direction of ($-\sin \theta$,$\cos \theta$) and
the slope for the entropy of the jamming power-law curve along this
direction ($-\sin \theta$,$\cos \theta$) is $0$ (see Fig.
\ref{slope}), that is,
\begin{equation}
 \frac {\partial S}{\partial \ln(\phi-\phi_{c})}\sin \theta = \frac {\partial S}{\partial \ln p}\cos \theta.
  \label{Eq_powerlog1}
\end{equation}


By the definition of angoricity $A=\partial \Gamma/\partial S$ and
compactivity $X=\partial V/\partial S$, we have:
\begin{equation}
 \frac {\partial S}{\partial \ln p} = p \frac {\partial S}{\partial p} = \Gamma \frac {\partial S}{\partial \Gamma} = \frac {\Gamma}{A} = \frac {c_1}{A},
 \label{Eq_powerlog2}
\end{equation}

\begin{equation}
\begin{split}
  \frac {\partial S}{\partial \ln(\phi-\phi_{c})} = & (\phi-\phi_{c}) \frac {\partial S}{\partial \phi}
  = (\phi-\phi_{c})\frac {\partial V}{\partial \phi}\frac {1}{X}= \\
  =&-(\phi-\phi_{c})\frac {N V_{g}}{\phi^{2}}\frac{1}{X}=-\frac {c_2}{X},
\end{split}
  \label{Eq_powerlog3}
\end{equation}
where $\phi=N V_{g}/V$, $c_1=\Gamma$ and $c_2=(\phi-\phi_{c})(N
V_{g}/\phi^{2})$.

By Eq. (\ref{Eq_powerlog2}) and Eq. (\ref{Eq_powerlog3}), we can
simplify Eq. (\ref{Eq_powerlog1}):
\begin{equation}
 \frac{c_1}{A}  + a \frac{c_2}{X} = 0.
\end{equation}

The relation between $X$ and $A$ can be obtained then
(Fig. \ref{slope}):
\begin{equation}
  X = -a \frac{c_2} {c_1} A=- a \frac{\phi-\phi_{c}}{p\phi}A.
  \label{Eq_powerlog4}
\end{equation}

From Eq. (\ref{Eq_powerlog4}) we obtain that: $X \propto
-(\phi-\phi_{c})^{1+a-\gamma}/\phi $ and near $\phi_{c}$:
\begin{equation}
  X \sim -(\phi-\phi_{c})^{2}.
\end{equation}

We notice that the compactivity is negative near the jamming
transition. A negative temperature is a general property of systems
with bounded energy like spins \cite{fluctuation}: the system attains
the larger volume (or energy in spins) at $\phi_c$ when $X\to 0^-$ and
not $X\to +\infty$ [The bounds $\phi_{c} \le \phi \le 1$ imply that
the jamming point at $X\to 0^-$ is ``hotter'' than $X\to +\infty$. At
the same time $A\to 0^+$ since the pressure vanishes].

We conclude that, $A$ and $X$ alone cannot play the role of
temperature, but a combination of both determined by entropy
maximization satisfying the coupling between stress and strain.
Instead, there is an actual ``jamming temperature'' $T_{\rm J}$ that
determines the direction $(-\sin \theta,\cos \theta)$ in the
$\log-\log$ plot of
Fig. \ref{entropydis} along the jamming equation of state (see
Fig. \ref{slope}).
By maximizing the entropy along this direction we obtain
the ``jamming temperature'' $T_{\rm J}$ as a function of $A$ and $X$:
\begin{equation}
  \frac {1}{T_{\rm J}} = \frac{c_1}{A}\sin \theta - \frac{c_2}{X} \cos \theta = \cos \theta (a \frac{c_1}{A} - \frac{c_2}{X}).
  \label{Eq_powerlog5}
\end{equation}

That is:

\begin{equation}
\begin{split}
  T_{\rm J} = &\frac{A \sin \theta}{c_1}=-\frac{X \cos \theta}{c_2} = \frac {\sin \theta}{\Gamma}A= \\
= & \frac{a}{\sqrt{1+a^2}} \frac{A}{\Gamma} \sim (\phi-\phi_{c})^{\gamma-a} \sim (\phi-\phi_{c}).
\end{split}
  \label{Eq_powerlog7}
\end{equation}
Thus, the temperature vanishes at the jamming transition.


Furthermore, the ``jamming energy'' $E_{\rm J}$, corresponding to
the ``jamming temperature'' $T_{\rm J}$ in Eq.
(\ref{Eq_powerlog5}), has the relation as below:
\begin{equation}
\begin{split}
   {\rm d}E_{\rm J} &=  T_{\rm J}{\rm d}S \\
                    &= T_{\rm J} \frac{\partial S}{\partial \ln(\phi-\phi_c)} {\rm d}\ln(\phi-\phi_c) + T_{\rm J} \frac{\partial S}{\partial \ln p} {\rm d}\ln p \\
                    &= (-\frac{X \cos \theta}{c_2})(-\frac {c_2}{X}) {\rm d}\ln(\phi-\phi_c) + \frac{A \sin \theta}{c_1} \frac {c_1}{A}{\rm d}\ln p \\
                    &= \cos \theta {\rm d}\ln(\phi-\phi_c) + \sin \theta {\rm d}\ln p \\
                    &=(\cos \theta + \sin \theta \tan \theta) {\rm d}\ln(\phi-\phi_c) \\
                    &=\frac {{\rm d}\ln(\phi-\phi_c)}{\cos \theta}.
  \label{Eq_powerlog9}
\end{split}
\end{equation}
That is,
\begin{equation}
 {\rm d}E_{\rm J} = \sqrt {a^2+1} {\rm d}\ln (\phi-\phi_c),
  \label{Eq_EJ2}
\end{equation}
and
\begin{equation}
 E_{\rm J} = (\sqrt {a^2+1}) \ln (\phi-\phi_c).
  \label{Eq_EJ3}
\end{equation}

By the definition of ``heat'' capacity, we obtain two jamming
capacities as the response to changes in $A$ and $X$:
\begin{equation}
\begin{array}{rl}
  C_{\rm \Gamma} \equiv &\partial \Gamma/\partial A \sim (\phi-\phi_{c})^{-1} \sim A^{-2/5}, \,\,\,\\
  C_{\rm V} \equiv &\partial V/\partial X \sim (\phi-\phi_{c})^{-1} \sim | X|^{-1/2}.
 \label{Eq_CX}
\end{array}
\end{equation}

The jamming capacity $C_{\rm J}$ can be obtained as:
\begin{equation}
  C_{\rm J} = T_{\rm J} \frac{\partial S}{\partial T_{\rm J}} = T_{\rm J}\frac{\partial S}{\partial \ln p}\frac{\partial \ln p}{\partial T_{\rm J}} + T_{\rm J}\frac{\partial S}{\partial \ln (\phi-\phi_{c})} \frac{\partial \ln (\phi-\phi_{c})}{\partial T_{\rm J}}.
  \label{Eq_powerlog6}
\end{equation}
Finally, with Eq. (\ref{Eq_powerlog1})--(\ref{Eq_powerlog3}), the capacity $C_{\rm J}$ can be calculated:
\begin{equation}
  C_{\rm J} = T_{\rm J} (\frac{c_1}{A}-\frac{c_2}{aX})\frac{\partial \ln p}{\partial T_{\rm J}} = T_{\rm J}\frac{1+a^2}{a^2}\frac{c_1}{A}\frac{\partial \ln p}{\partial T_{\rm J}}.
  \label{Eq_powerlog8}
\end{equation}
Since $T_{\rm J} \sim (\phi-\phi_{c})$ and $p \sim (\phi-\phi_{c})^{1.5}$,
we obtain
\begin{equation}
C_{\rm J} \sim (\phi-\phi_{c})^{-1}.
\end{equation}

From Eq. (\ref{Eq_CX}), the jamming capacities diverge at the jamming
transition as $A \to 0^+$ and $X \to 0^{-}$. However, this result does
not imply that the transition is critical since from fluctuation
theory of pressure and volume \cite{fluctuation} we obtain:
\begin{equation}
\begin{split}
  \langle (\Delta \Gamma)^{2} \rangle = A^{2}C_{\rm \Gamma} \sim A^{1.6},\\
 \langle (\Delta V)^{2} \rangle = X^{2}C_{\rm V} \sim |X|^{1.5}.
\end{split}
 \label{Eq_DV}
\end{equation}
Thus, the pressure and volume fluctuations near the jamming transition
do not diverge, but instead vanish when $A \rightarrow 0^+$ and $X
\rightarrow 0^{-}$.  From a thermodynamical point of view, the
transition is not of second order due to the lack of critical
fluctuations.  As a consequence, no diverging static correlation
length from a correlation function can be found at the jamming
point. However, other correlation lengths of dynamic origin may still
exist in the response of the jammed system to perturbations, such as
those imposed by a shear strain or in vibrating modes
\cite{mode1,mode2}. Such a dynamic correlation length would not appear
in a purely thermodynamic static treatment as developed here. We note
that static anisotropic packings can be treated in the present
formalism by allowing the inverse angoricity to be tensorial
\cite{canonicaljam}.

The intensive jamming temperature Eq. (\ref{Eq_powerlog7}) gives use to a
jamming effective energy $E_{\rm J}$ as the extensive variable
satisfying $T_{\rm J}=\partial E_{\rm J}/\partial S$ and a full
jamming capacity $C_{\rm J} \sim (\phi-\phi_{c})^{-1}$, which also
diverges at jamming. However, the fluctuations of $E_{\rm J}$ defined as
$\langle (\Delta E_{\rm J})^{2} \rangle = T_{\rm J}^{2}C_{\rm J} \sim
T_{\rm J}$ has the same behavior as the fluctuations of volume and
pressure, vanishing at the jamming transition $T_{\rm J}\to 0^+$ [$A
\to 0^+$ in Eq. (\ref{Eq_powerlog7})].

\section{Comparison with O'Hern et al.}
\label{comparison}

The results so far show a general agreement between MD and the
ensemble average. These include the maximum entropy principle and
ergodicity. We now turn to a comparison with similar simulations done
by O'Hern {\it et al.} \cite{ohern,manypackings}. These studies
perform an exhaustive search of all configurations in the PEL of
frictionless particles similarly as in the present paper. However,
they find that the microstates are not equiprobable, i.e., microstates
with the same pressure and volume fraction (pressure is fixed at zero
since only hard sphere states are of interest) do not have the same
probability when sampled by a given algorithm. Furthermore,
experimental studies of equilibration between two systems
\cite{leche}, suggests that a hidden variable is necessary to describe
the microstates, further supporting the results of \cite{ohern}. The
applicability of the microcanonical ensemble is based on the fact that
the microstates are defined by $(\Gamma,\phi)$. Thus, the fact that
the states are not equiprobable implies that there must be an extra
variable needed to describe their probabilities. Therefore, ergodicity
and the maximum entropy principle, which are downstream from
equiprobability, are not supposed to hold,
in disagreement with the results shown in the present paper.

To investigate this situation, we repeat the same calculations as in
\cite{ohern} with our algorithms. We first rule out subtleties related
to algorithmic dependent results in sampling the space of
configurations. We use our 30 particles system and use $\phi=0.61$
very close to jamming and $\Gamma=0$ to look for the hard sphere
packings.  We search for the jammed configurations as above. We recall
that the sampling of the space of configurations is not complete due
to the relatively large system size but represent a good sampling as
discussed above. Ref. \cite{ohern} uses a different system of 14
particles in 2d for which 248,900 configurations are found
exhaustively sampling the phase space (which is estimated to have
$\sim 371,500$ states). These simulations correspond to a system with
periodic boundary conditions for which a larger space is expected than
the close boundary-system of Section \ref{example}.  However, these
differences do not affect the conclusions below.

\begin{figure}
  \centerline{ \resizebox{7.0cm}{!} {
    \includegraphics{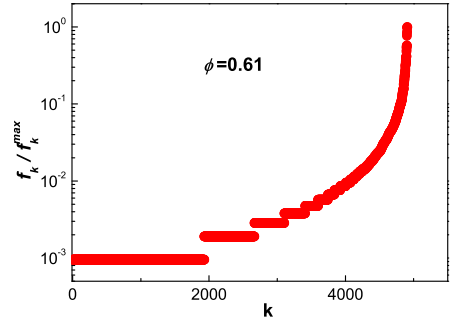}}}
\caption{Sampling probability of each microstate $f_k$ identified by
  its rank $k$ fro low to high. Results are for a system of 30
  particles at $\phi=0.61$ and a narrow set of pressures around 0.}
\label{kun1}
\end{figure}

We start by measuring $f_k$ which is the probability to find a given
microstate $k$ as defined by \cite{ohern}: each packing can be
obtained many times during a search and therefore $f_k$ measures the
probability for which each packing occurs. The main result of
\cite{ohern} is that $f_k$ differs by many orders of magnitude for
states with fixed $(\Gamma,\phi)$. Indeed, even configurations which
are visually very similar can be $10^6$ more frequent, see Fig. 1 of
\cite{ohern}.

Figure \ref{kun1} shows $f_k$ sorted as a function of $k$, the rank,
as in \cite{ohern}.  This plot reproduces the results of \cite{ohern}
in our system. For a fixed pressure and volume there are many states
with a large difference in their probability. The least probable
states are 10$^{-3}$ less probable than the most probable state
showing a breakdown of equiprobability. The question is how to
interpret the results of ergodicity in the light of the failure of
equiprobability and whether there is a need for an extra variable to
describe the microstates.

We first mention the issue of the small system size. It is quite
possible that the low probability states will completely disappear in
the thermodynamic limit and the ones remaining are the most probable ones
with equal probability. Indeed, the flat average assumption is only
valid in the thermodynamic limit and simply says that even if there
exists less probable states ($10^{-3}$ less probable) then they will
be irrelevant in the ensemble average, thus only the most probable and
flat states are important.
\begin{figure}
  \centerline{ (a) \resizebox{7.0cm}{!} {
    \includegraphics{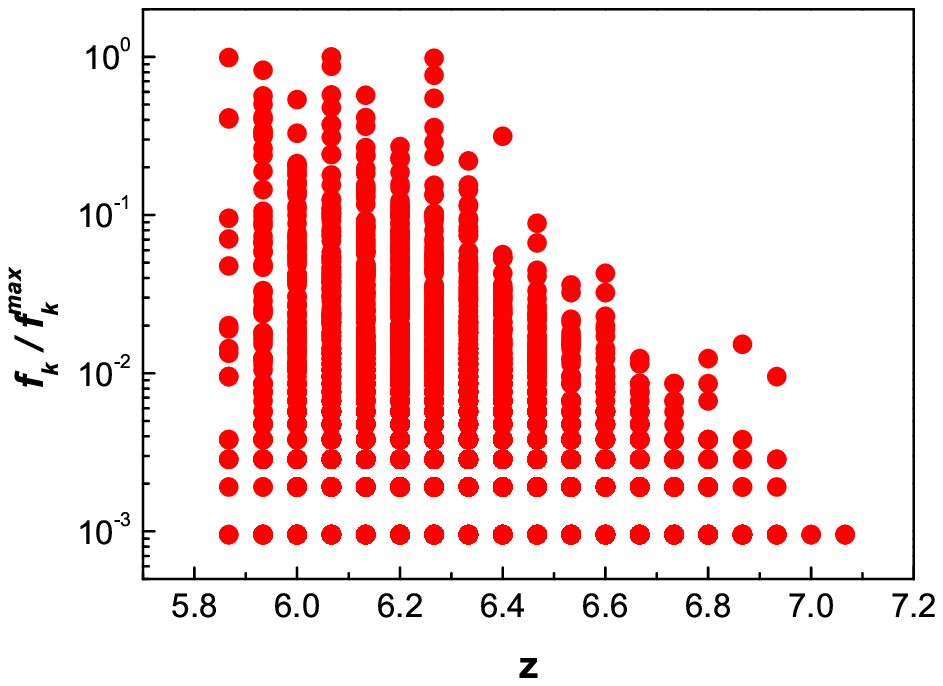}}}
  \centerline{ (b) \resizebox{7.0cm}{!} {
    \includegraphics{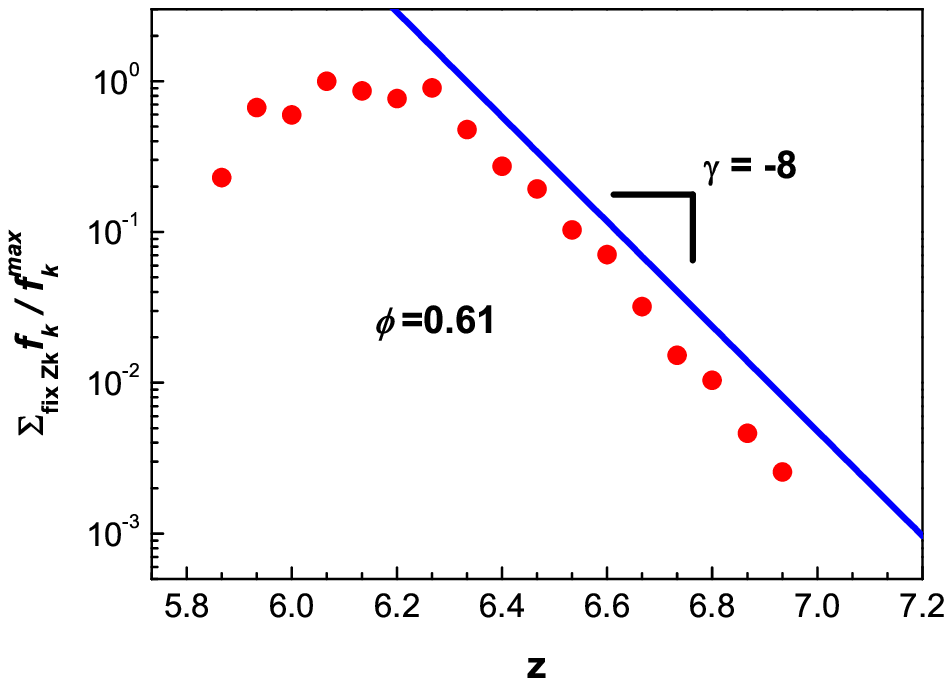}}}
\caption{ (a)
  Sampling probability of each microstate $f_k$ as a function of the
  coordination number $Z_k$ of each microstate. (b) Plot of $\ln
  (\sum_{{\rm fix} Z_k} f_k/f_k^{max})$ versus $Z_k$ showing an
  exponential decay consistent with the density of states proposed in
  \cite{compactivity}.}
\label{kun2}
\end{figure}

We have done simulations with $N=$14 particles and found that the
least probable states are $10^{-5}$ less probable than the most
probable states. Comparing with the factor $10^{-3}$ for $N=30$, may
indicate that the system size may take care of the non-equiprobability
problem. However, calculations for larger system to fully test this
assertion are out of the range of current and near future
computational power.

Second, we notice that the coordination number is also important to
define the jammed states.  Figure \ref{kun2} plots the same states as
Fig. \ref{kun1} but as a function of $Z_k$, the coordination number of
microstate $k$.  The most probable states satisfy:
\begin{equation}
f_k(Z_k) \sim e^{-8 Z_k}.
\label{fk}
\end{equation}

Furthermore, if we sum up all the states for a given $Z_k$ and plot
$\log(\sum_{{\rm fix} Z_k} f_k)$ vs $Z_k$ we obtain Eq. (\ref{fk}) as
seen in Fig.  \ref{kun2}b. This result does not mean that $Z_k$ is the
hidden variable but rather Eq. (\ref{fk}) provides the density of
states proposed in \cite{compactivity} in the thermodynamics
calculation of the random close packing of spheres. Indeed, we have
predicted that the density of states $g(z) = h_z^z$, with $h_z$
playing the role of a Planck constant defining the minimum size in the
volume landscape. According to Eq. (\ref{fk}), this prediction is
satisfied in average with $h_z=e^{-8}$ which is a small number as
expected.

This result indicates that some variability in the probabilities of the
microstates is expected from the fluctuations in the coordination
number of each microstate. In Appendix \ref{z} we elaborate an
extension of the framework of \cite{compactivity} to incorporate
fluctuations in $Z$ that are neglected in \cite{compactivity}. The
purpose is to test whether the RCP and jamming transition are affected
by these fluctuations. We find that the results are consistent with
those found in \cite{compactivity}.

We notice that for a fix $Z_k$ there are still many marginal states
with very small probabilities as seen in Fig. \ref{kun2}a. If these
states do not completely disappear in the thermodynamic limit, then
they need to be explained. We end this discussion by providing a
possible explanation for the existence of these states.

The numerical breakdown of equiprobability might be related to the
fact that the found packings are not indistinguishable. Indeed, we
ignore the rotation and translation symmetries of the packing in order
to make the numerical search possible. However, for the Edwards flat
hypothesis, these packings should be assumed different. Once we
breakdown the rotational symmetry, there would be many similar
packings. The high degeneracy of the high symmetric packings may be
responsible for the uneven distribution, which would be, in this case,
simply artificial.

For instance, consider two packings with 4 particles: (a) a square
packing with each particle on the corner and (b) a triangle with each
particle in each corner plus one in the center.

For both packings there are 4! = 24 different permutations, which
should be considered as 24 different packings, in principle. However,
since we can rotate the square packing by 90 degree and obtain the
same one, there are only 24/4 = 6 distinguishable packings. Similarly,
for the triangle, there are 24/3 = 8 distinguishable packings. The
probability between (a) and (b) is uneven (6:8) if we assume that each
distinguishable packing is equal-probable. Therefore, different
symmetries of the packings may contribute to the unequal probabilities
that we measure in the algorithms.

Therefore, if the Edwards assumption is correct, $f_k$ should be
proportional to $S_k$, where $S_k$ is the order of the symmetry group
(point group) of the packing $k$, since there are $S_k$ degenerations
(same packing if particles are identical).  This conjecture needs
extra evaluation of the symmetry of each packing.  For instance, the
translation invariance is important, and for cubic periodic boundary,
it is also important to include the symmetry of cubic point group
C$_{\rm 3h}$.

We do not investigate this conjecture but rather provide the codes and
packings in http://jamlab.org to do that.  Since the 3d case is
complicated, one might try the 2d system first to easily visualize
different packings. A simple question is: given two packings with
different frequencies, how do they look like \cite{ohern}?  Would be
the high symmetric one visited more, or inversely?

\section{Conclusion }
\label{conclusion}

We have demonstrated that the concept of `` thermalization '' at a
compactivity and angoricity in jammed systems is reasonable by the
direct test of ergodicity. The numerical results indicate that
the full canonical ensemble of pressure and volume describes the
observables near the jamming transition quite well.
From a static thermodynamic viewpoint, the jamming phase transition
does not present critical fluctuations characteristic of second-order
transitions since the fluctuations of several observables vanish
approaching jamming.  The lack of critical fluctuations is respect to
the angoricity and compactivity in the jammed phase $\phi \to
\phi_{c}^+$, which does not preclude the existence of critical
fluctuations when accounting for the full range of fluctuations in the
liquid to jammed transition below $\phi_c$. Thus, a critical diverging
length scale might still appear as $\phi \to \phi_c^-$ \cite{glass0},
which has been recently observed by experiment \cite{glass1}.

In
conclusion, our results suggest an ensemble treatment of the jamming
transition. One possible analytical route to use this formalism would
be to incorporate the coupling between volume and coordination number
at the particle level found in \cite{compactivity,j1} together with
similar dependence for the stress to solve the partition
function. This treatment would allow analytical solutions for the
observables with the goal of characterizing the scaling laws near the
jamming transition.




Acknowledgements: We thank NSF-CMMT and DOE-Geosciences Division for
financial support and L. Gallos for discussions.

\clearpage

\appendix

\section{Microstates and Fluctuations in coordination number}
\label{z}

Here, we develop a $Z$-ensemble for hard spheres in the limit of zero
angoricity. In the main test we found that fluctuations in $Z$ may
account for certain variability in the probability of
microstates. Here we investigate whether this variability affect the
existence of RCP and the jamming point. We develop a partition
function in Edwards ensemble to study the dependence of RCP on this
type of fluctuations.

The partition function is
\begin{equation}
\mathcal{Z} = \int\ldots\int_{Nz_{min} < \sum z_i < Nz_{max}}
\prod_i e^{-(z_i/z^* + \beta\kappa/z_i)}dz_i,
\end{equation}
where $z_{min} = Z$ and $z_{max}=6$, $\beta=1/X$, and
$\kappa=2\sqrt{3}$. We follow the notation and concepts from
\cite{compactivity,j1,j2,j3}.  We define $x = (\sum_i z_i)/N$, thus:
\begin{equation}
\mathcal{Z} = \int_{z_{min}}^{z_{max}}P(x)dx,
\end{equation}
where
\begin{equation}
  P(x) \equiv \int_{0}^{\infty}\ldots\int_{0}^{\infty}\prod_i
  e^{-(z_i/z^* + \beta\kappa/z_i)} \delta\left(x - \frac{1}{N}\sum_i
    z_i\right)dz_i,
\end{equation}
where $z^*=1/8$ according to Fig. \ref{kun2}b. We  consider the inverse
Fourier transform of $P_x(f)$:
\begin{widetext}
\begin{equation}
\begin{split}
\mathcal{F}_f^{-1}[P_x(f)]\equiv \int_{-\infty}^{\infty} e^{2\pi
i f X} P(x) dx
= \int_{0}^{\infty}\ldots\int_{0}^{\infty}\prod_i e^{-(z_i/z^* +
\beta\kappa/z_i)} e^{2\pi i f \sum z_i/N}dz_i = \\
 = \left[\int_{0}^{\infty}e^{-(z/z^* + \beta\kappa/z)} e^{2\pi i
f z/N}dz\right]^N
=\left\{\int_{0}^{\infty} \left[ 1+\left(\frac{2\pi
ifz}{N}\right)+\frac{1}{2}\left(\frac{2\pi ifz}{N}\right)^2 +
\ldots \right]e^{-(z/z^* + \beta\kappa/z) } \right\}^N.
\end{split}
\end{equation}
\end{widetext}
Since
\begin{equation}
\int_0^{\infty} x^n e^{-\frac{a}{2}(x+1/x)} dx = 2 K_n(a),
\end{equation}
where $K_n(a)$ is the modified Bessel function of the second kind.
By taking the coupling constant
\begin{equation}
\begin{split}
B \equiv \beta\kappa/z^*, \\
a
\equiv 2B^{1/2}, \\
z = B^{1/2} z^* x.
\end{split}
\end{equation}
Then:
\begin{equation}
\int_0^{\infty} z^n e^{-(z/z^* + \beta\kappa/z)} dz = 2 {z^*}
^{n+1} B^{(n+1)/2} K_n(2 B^{1/2}).
\end{equation}

\begin{widetext}
Thus,
\begin{equation}
\begin{split}
\mathcal{F}_f^{-1}[P_X(f)] &= (2z^*)^N\left[B^{1/2}K_0(2 B^{1/2})
+ \left(\frac{2\pi ifz^* }{N}\right) B K_1(2 B^{1/2}) +
\frac{1}{2}\left(\frac{2\pi ifz^* }{N}\right)^2 B^{3/2} K_2(2
B^{1/2})+
O(N^{-3})\right]^N\\
&=(2z^*)^N\exp \left\{ N\ln\left[B^{1/2}K_0(2 B^{1/2}) +
\left(\frac{2\pi ifz^* }{N}\right) B K_1(2 B^{1/2}) + \frac{1}{2}
\left(\frac{2\pi ifz^* }{N}\right)^2 B^{3/2} K_2(2 B^{1/2})+
O(N^{-3})\right]\right\}\\
&=(2z^* B^{1/2}K_0(2 B^{1/2}))^N\exp \left\{ N\ln\left[1 +
\left(\frac{2\pi ifz^* }{N}\right) \frac{K_1(2 B^{1/2})}{K_0(2
B^{1/2})}B^{1/2}  + \frac{1}{2}\left(\frac{2\pi ifz^*
}{N}\right)^2 \frac{K_2(2 B^{1/2})}{K_0(2 B^{1/2})}B+
O(N^{-3})\right]\right\}
\end{split}
\end{equation}
Now, we expand
\begin{equation}
\ln(1+x) = x - \frac{1}{2}x^2 + \frac{1}{3} x^3 + \ldots
\end{equation}

and

\begin{equation}
\begin{split}
&\exp \left\{ N\ln\left[1 + \left(\frac{2\pi ifz^* }{N}\right)
\frac{K_1(2 B^{1/2})}{K_0(2 B^{1/2})}B^{1/2}  +
\frac{1}{2}\left(\frac{2\pi ifz^* }{N}\right)^2 \frac{K_2(2
B^{1/2})}{K_0(2 B^{1/2})}B+
O(N^{-3})\right]\right\} \\
=&\exp \left\{ N\left[\left(\frac{2\pi ifz^* }{N}\right)
\frac{K_1(2 B^{1/2})}{K_0(2 B^{1/2})}B^{1/2}  +
\frac{1}{2}\left(\frac{2\pi ifz^* }{N}\right)^2 \frac{K_2(2
B^{1/2})}{K_0(2 B^{1/2})}B- \frac{1}{2}\left(\frac{2\pi ifz^*
}{N}\frac{K_1(2 B^{1/2})}{K_0(2
B^{1/2})}\right)^2 B+ O(N^{-3})\right]\right\}\\
\approx & \exp\left[ 2\pi if \left(z^*B^{1/2}\frac{K_1(2
B^{1/2})}{K_0(2 B^{1/2})}\right)
 - \frac{(2\pi f)^2}{2N} {z^*} ^{2} B\left(\frac{K_2(2
B^{1/2})}{K_0(2 B^{1/2})} - \frac{K_1(2 B^{1/2})^2}{K_0(2
B^{1/2})^2}\right)\right],
\end{split}
\end{equation}

\end{widetext}
is just a Gaussian distribution with the mean
\begin{equation}
\mu = z^*B^{1/2}\frac{K_1(2 B^{1/2})}{K_0(2 B^{1/2})},
\end{equation}
and the mean square deviation
\begin{equation}
\sigma_{N} = \frac{\sigma}{\sqrt{N}},
\end{equation}
where
\begin{equation}
\sigma^2 \equiv {z^*} ^{2} B\left(\frac{K_2(2 B^{1/2})}{K_0(2
B^{1/2})} - \frac{K_1(2 B^{1/2})^2}{K_0(2 B^{1/2})^2}\right).
\end{equation}

Thus, by using the saddle point approximation, we obtain the free
energy density $f$:
\begin{equation}
\begin{split}
  \beta f \equiv& -\lim_{N \rightarrow \infty}
  \frac{\ln(\mathcal{Z})}{N} = \\
  -\ln(B^{1/2}K_0(2 B^{1/2})) +& \frac{1}{2\sigma^2}[ (\mu -
    z_{max})^2\Theta(\mu - z_{max}) +
    \\
    & (z_{min}-\mu)^2\Theta(z_{min}-\mu)].
\end{split}
\end{equation}

We also obtain the energy density, or volume density in the context of
Edwards:

\begin{equation}
\begin{split}
\frac{z^*}{\kappa} w = \frac{d(\beta f)}{d B} = -\frac{1}{2B} +
B^{-1/2}\frac{K_1(2 B^{1/2})}{K_0(2 B^{1/2})} +\\
\frac{1}{2}\frac{d}{dB} \left[ \frac{(\mu - z_{max})^2}{\sigma^2}
\right] \Theta(\mu - z_{max}) +\\ \frac{1}{2}\frac{d}{dB}
\left[\frac{(z_{min}-\mu)^2}{\sigma^2}\right]\Theta(z_{min}-\mu).
\end{split}
\end{equation}

\begin{equation}
\frac{(\mu - z_{max})^2}{\sigma^2} =
\frac{(L(B)-Z_{max})^2}{B+L(B)-L(B)^2},
\end{equation}
\begin{equation}
\frac{(\mu - z_{min})^2}{\sigma^2} =
\frac{(L(B)-Z_{min})^2}{B+L(B)-L(B)^2},
\end{equation}
where $ Z_{max} \equiv z_{max}/z^*$, $ Z_{min} \equiv z_{min}/z^*,$
and
\begin{equation}
L(B) \equiv B^{1/2}\frac{K_1(2 B^{1/2})}{K_0(2 B^{1/2})},
\end{equation}
because
\begin{equation}
\frac{d L(B)}{d B} = \frac{L(B)^2}{B}-1.
\end{equation}

\begin{widetext}

Then,
\begin{equation}
  \frac{B}{2}\frac{d}{dB} \left[ \frac{(\mu - z_{max})^2}{\sigma^2}
  \right] \Theta(\mu - z_{max}) =
  \left[\frac{1}{2}\left(\frac{L(B)(L(B)-Z_{max})}{B+L(B)-L(B)^2}\right)^2-\frac{(B-Z_{max} L(B))(L(B)-Z_{max})}{B+L(B)-L(B)^2}
  \right]\Theta(L(B)-Z_{max}),
\end{equation}

and

\begin{equation}
\frac{B}{2}\frac{d}{dB} \left[ \frac{(\mu - z_{min})^2}{\sigma^2}
\right] \Theta(\mu - z_{min}) =
\left[\frac{1}{2}\left(\frac{L(B)(L(B)-Z_{min})}{B+L(B)-L(B)^2}\right)^2-
\frac{(B-Z_{min} L(B))(L(B)-Z_{min})}{B+L(B)-L(B)^2}
\right]\Theta(Z_{min}-L(B)).
\end{equation}

Thus,
\begin{equation}
\begin{split}
\beta w = -\frac{1}{2} + L(B) +
\left[\frac{1}{2}\left(\frac{L(B)(L(B)-Z_{max})}{B+L(B)-L(B)^2}\right)^2
-\frac{(B-Z_{max} L(B))(L(B)-Z_{max})}{B+L(B)-L(B)^2}
\right]\Theta(L(B)-Z_{max})
\\
+\left[\frac{1}{2}\left(\frac{L(B)(L(B)-Z_{min})}{B+L(B)-L(B)^2}\right)^2
-\frac{(B-Z_{min} L(B))(L(B)-Z_{min})}{B+L(B)-L(B)^2}
\right]\Theta(Z_{min}-L(B)),
\end{split}
\end{equation}
\end{widetext}

and the entropy density:
\begin{equation}
\begin{split}
s = \beta (w-f).
\end{split}
\end{equation}

There are two phase transitions at $L(B) = Z_{min}$ and $L(B) =
Z_{max}$. For the jammed phase $Z_{min} < L(B) < Z_{max}$, we have
$\beta w = L(B)-1/2$. If $z^*$ is a small value, $z^*=1/8$ from
Fig. \ref{kun2}, then $B$ is relatively large. Thus, $L(B) \approx
B^{1/2}$ and $w_{max} \approx L(B)/\beta = \kappa / z^* B^{-1/2} =
\kappa/ (z^* Z_{min}) = \kappa/z_{min}$. Similarly, $w_{min} \approx
\kappa/z_{max}$, which is consistent with the boundaries of the phase
diagram obtained in \cite{compactivity}.  Furthermore, $f \approx
2B^{1/2}$ and $s \approx s_0-B^{1/2}$, where $s_0 = Z_{max}$. Or, $s =
(z_{max} - \kappa/w)/z^*$. Thus, we have verified that the inclusion
of fluctuations in the coordination number does not change the shape
of the jamming phase diagram obtained in \cite{compactivity,j2}. These
fluctuations may affect the probability of the microstates according
to the density of states proposed in \cite{compactivity}. A further
application of this generalized $Z$-ensemble is developed in \cite{j4}
to calculate the probability of coordination numbers in packings, with
good agreement with the numerical results for different packings in
the phase diagram.


\begin{thebibliography}{99}
\bibitem{clarification} It is worth the clarification. In the study of
  \cite{powerlaw} a different exponent $b\ne 1/2$ in frictionless
  packings was reported.  The reason might be due to the protocol
  employed which uses a servo mechanism \cite{hmlaw2} to constantly
  adjust the strain to achieved a predetermined pressure. We think
  that at unrealistic low pressures (below 100 KPa) the protocol may
  not equilibrate the packings properly for the parameters used. In
  fact, there is an excess of small forces (noticeable in the force
  distribution which shows a full exponential rather than the plateau
  at low forces) probably arising from some particles which were not
  fully equilibrated. In a later study \cite{forcebalance2} a
  different protocol was used, called split algorithm in
  \cite{compactivity}, with the result $b=1/2$.
\end{thebibliography}
\end{document}